\documentclass[journal,subeqn,caption]{IEEEtran}
\usepackage{mathtools}
\usepackage{algorithm}
\usepackage{algpseudocode}
\usepackage{hhline}
\usepackage{pdflscape}
\usepackage{balance}

\usepackage{cite}
\usepackage{float}
\usepackage{multicol}
\usepackage{multirow}
\usepackage{balance}
\usepackage{graphicx}
\usepackage{mathdots}

\usepackage{rotating,booktabs,dcolumn}
\usepackage[flushleft]{threeparttable}

\DeclarePairedDelimiter\abs{\lvert}{\rvert}%
\DeclarePairedDelimiter\norm{\lVert}{\rVert}%

\makeatletter
\let\oldabs\abs
\def\abs{\@ifstar{\oldabs}{\oldabs*}}
\let\oldnorm\norm
\def\norm{\@ifstar{\oldnorm}{\oldnorm*}}
\makeatother

\algdef{SE}[SUBALG]{Indent}{EndIndent}{}{\algorithmicend\ }%
\algtext*{Indent}
\algtext*{EndIndent}

\usepackage{array}
\newcolumntype{P}[1]{>{\centering\arraybackslash}p{#1}}
\newcolumntype{M}[1]{>{\centering\arraybackslash}m{#1}}

\usepackage{hyperref}
\usepackage{graphicx}
\usepackage{amsfonts}
\usepackage{amsmath}
\usepackage{mathabx}
\usepackage{bm}
\usepackage{color}
\usepackage{epstopdf}
\usepackage{mathtools, nccmath}
\usepackage{multirow}
\usepackage{bbm}
\usepackage{marginnote}
\usepackage{tikz}
\usetikzlibrary{shapes,arrows,chains}
\usepackage{enumitem}
\usepackage{comment}
\usepackage{eurosym}
 \usepackage{epstopdf}
\usetikzlibrary{arrows,decorations.pathmorphing,backgrounds,fit,positioning,shapes.symbols,chains}
\usepackage{caption}
\usepackage{colortbl}
\usepackage{float}
\usepackage{xcolor}
\usepackage{subfig}
\usepackage{tabularx}

\DeclareCaptionLabelSeparator{dotspace}{.\hspace*{1em}}
\definecolor{Gray}{gray}{0.9}
\definecolor{LightCyan}{rgb}{0.88,1,1}

\newcolumntype{a}{>{\columncolor{Gray}}c}
\newcolumntype{b}{>{\columncolor{white}}c}

\makeatletter
\newcases{Rcases}{$\mskip\thickmuskip$}{%
  \hfil$\m@th{##}$}{$\m@th{##}$\hfil}{.}{\rbrace}
\makeatother
\allowdisplaybreaks

 \ifCLASSINFOpdf
\else
\fi
\newenvironment{ldescription}[1]
  {\begin{list}{}%
   {\renewcommand\makelabel[1]{##1\hfill}%
   \settowidth\labelwidth{\makelabel{#1}}%
   \setlength\leftmargin{\labelwidth}
   \addtolength\leftmargin{\labelsep}}}
  {\end{list}}

\begin{document}

\newcommand{\Xiul}{\Xi^{\mathrm{ul}}}
\newcommand{\Xill}{\Xi^{\mathrm{ll}}}

\newcommand{\SoEmax}{\overline{\bm{SoE}}}
\newcommand{\etach}{\bm{\eta}^{\mathbf{ch}}}
\newcommand{\etadis}{\bm{\eta}^{\mathbf{dis}}}
\newcommand{\qchmax}{\bm{\overline{q}}^{\mathbf{ch}}}
\newcommand{\qdismax}{\bm{\overline{q}}^{\mathbf{dis}}}

\newcommand{\SoE}{SoE_{t}}
\newcommand{\SoEm}{SoE_{t-1}}
\newcommand{\qbat}{p^{\mathrm{ES}}_{t}}
\newcommand{\qch}{p^{\mathrm{ch}}_{t}}
\newcommand{\qdis}{p^{\mathrm{dis}}_{t}}

\newcommand{\qchi}{p^{\mathrm{ch}}_{t,i}}
\newcommand{\qdisi}{p^{\mathrm{dis}}_{t,i}}

\newcommand{\compprofitin}{\mathcal{C}_{i,n}}
\newcommand{\veriprofitin}{\mathcal{V}_{i,n}}
\newcommand{\veriprofitimean}{\overline{\mathcal{V}_{i}}}

\newcommand{\qbatcenter}{{p^{\mathrm{ES, cnt}}_{t, i}}}
\newcommand{\qbatcenternext}{{p^{\mathrm{ES, cnt}}_{t, i+1}}}
\newcommand{\qbatrad}{{p^{\mathrm{ES, rad}}_{t, i}}}
\newcommand{\qbatradnext}{{p^{\mathrm{ES, rad}}_{t, i+1}}}
\newcommand{\qbatopt}{{p^{\mathrm{ES, opt}}_{t}}}
\newcommand{\qbati}{p^{\mathrm{ES}}_{t,i}}
\newcommand{\qbatimean}{\overline{p^{\mathrm{ES}}}_{t,i}}
\newcommand{\qbatin}{p^{\mathrm{ES}}_{t,i,n}}
\newcommand{\qbatvect}{\left(p^{\mathrm{ES}}_1, p^{\mathrm{ES}}_2, \ldots, p^{\mathrm{ES}}_{|\tau|}\right)}
\newcommand{\qbativect}{\left(p^{\mathrm{ES}}_{1,i}, p^{\mathrm{ES}}_{2,i}, \ldots, p^{\mathrm{ES}}_{|\tau|,i}\right)}
\newcommand{\qbatinvect}{\left(p^{\mathrm{ES}}_{1,i,n}, p^{\mathrm{ES}}_{2,i,n}, \ldots, p^{\mathrm{ES}}_{|\tau|,i,n}\right)}
\newcommand{\qbatimeanvect}{\left(\overline{p^{\mathrm{ES}}_{1,i}}, \overline{p^{\mathrm{ES}}_{2,i}}, \ldots, \overline{p^{\mathrm{ES}}_{|\tau|,i}}\right)}

\newcommand{\xch}{x^{\mathrm{ch}}_{t}}
\newcommand{\xchi}{x^{\mathrm{ch}}_{t,i}}

\newcommand{\lambdaP}{\lambda_{t}}
\newcommand{\profit}{\Omega}

\newcommand{\eF}{F}
\newcommand{\eFapprox}{\hat{F}}


\title{Solving Bilevel Optimal Bidding Problems Using Deep Convolutional Neural Networks}
\author{D. Vlah, K. \v{S}epetanc, \textit{Student Member}, \textit{IEEE} and H. Pand\v{z}i\'c, \textit{Senior Member}, \textit{IEEE}
\thanks{The authors are with the University of Zagreb Faculty of Electrical Engineering and Computing -- UNIZG-FER (emails: domagoj.vlah@fer.hr; karlo.sepetanc@fer.hr; hrvoje.pandzic@fer.hr) and the Innovation Centre Nikola Tesla (ICENT). 
\newline
Employment of Karlo \v{S}epetanc at UNIZG-FER is fully funded by the Croatian Science Foundation under programme DOK-2018-09. 
The research leading to these results has received funding from the European Union’s Horizon 2020 research and innovation programme under grant agreement No 863876 (project FLEXGRID). The sole responsibility for the content of this document lies with the authors. It does not necessarily reflect the opinion of the Innovation and Networks Executive Agency (INEA) or the European Commission (EC). INEA or the EC are not responsible for any use that may be made of the information contained therein.}
%
%
}
%
%
\maketitle

\begin{abstract}
Current state-of-the-art solution techniques for solving bilevel optimization problems either assume strong problem regularity criteria or are computationally intractable. In this paper we address power system problems of bilevel structure, commonly arising after the deregulation of the power industry. Such problems are predominantly solved by converting the lower-level problem into a set of equivalent constraints using the Karush-Kuhn-Tucker optimality conditions at an expense of binary variables. Furthermore, in case the lower-level problem is nonconvex, the strong duality does not hold rendering the single-level reduction techniques inapplicable. To overcome this, we propose an effective numerical scheme based on bypassing the lower level completely using an approximation function that replicates the relevant lower level effect on the upper level. The approximation function is constructed by training a deep convolutional neural network. The numerical procedure is run iteratively to enhance the accuracy. 

As a case study, the proposed method is applied to a price-maker energy storage optimal bidding problem that considers an AC power flow-based market clearing in the lower level. The results indicate that greater actual profits are achieved as compared to the less accurate DC market representation.

\end{abstract}

\section{Introduction}\label{sec:intro}
\subsection{Background and paper scope}

Deregulation and liberalization of the power sector worldwide dislodged large monopolistic power utilities, allowing for private companies to become important players in the sector. However, each of the newly created entities have their own goal, e.g. generating companies want to maximize their profit, system operators maximize the security of supply, while market operators maximize social welfare. Because of the increased number of players with conflicting objectives, they need to consider each other's goals and objective functions when optimizing their own utility. To accommodate the interaction between own and other player's actions, the researchers commonly resort to bilevel models, where own optimization problem (the upper-level (UL) problem) is constrained by another optimization problem (the lower-level (LL) problem). This setting assumes that the lower-level behavior is known, which is the case when considering the market operator conducting its market-clearing procedure or any other regulated entity that behaves according to some widely-known rules.

As of October 5, 2021, the IEEE Xplore database \cite{ieee} indicates that three flagship IEEE Power and Energy Society journals published 182 journals with word \emph{bilevel} in the title (115 such papers in IEEE Transactions on Power Systems, 45 in IEEE Transactions on Smart Grid and 22 in IEEE Transactions on Sustainable Energy). These papers cover a wide range of bilevel problems. Some of the most common topics include protection of a power system against a terrorist attack, e.g. \cite{terror}, pricing schemes, e.g. \cite{pricing}, maintenance scheduling, e.g. \cite{maintain}, expansion planning, e.g. \cite{TEP}, or optimal bidding in one or more energy \cite{energy_m} or financial markets \cite{fin_m}.

Although bilevel models have been used extensively in the literature, they often suffer from two drawbacks. The first one is related to linearization, as commonly one or more variables from the lower-level problem appear multiply an upper-level variable in the upper-level objective function. Although in many cases this can be linearized using the strong duality theorem and some of the Karush-Kuhn-Tucker (KKT) optimality conditions \cite{KKT}, see e.g. \cite{HPmaintain}, in some cases this is not possible. In such cases the authors commonly resort to the binary expansion method (see appendix B in \cite{energy_m}). However, besides being an approximation, this method can result in intolerable computational times, bringing us to the second drawback, i.e. computational (in)tractability. Issues with tractability often arise when the lower-level problem is stochastic or has many inequality constraints, resulting in a large number of binary variables. Some authors thus resort to an iterative procedure that considers the complicating dual variables in the problematic bilinear terms as parameters, and updating their values in the following iteration \cite{param}.

The aim of this paper is to present a numerical scheme based on deep convolutional neural networks paired with state-of-the-art training procedure for solving complex bilevel problems arising in the power systems community. As a representative of such problems, we solve a bilevel problem of optimal participation of an energy storage in the day-ahead energy market. We assume an AC-optimal power flow (OPF)-based market clearing algorithm in the lower-level. AC OPF is a challenging problem with numerous simplification attempts, e.g. by convexification \cite{cpsota}. However, to this date there is still no known exact finite convex AC OPF formulation that classical approaches could solve to optimality.

Optimal bidding problems consist of two interlinked optimizations. The first problem, also called the leader or the upper-level problem, represents the market participant that maximizes the agent's profit due to arbitrage, while the second problem, also called the follower or the lower-level problem, is the market clearing that maximizes the social welfare and determines the electricity prices that depend on the bids from the upper-level problem. The described bilevel optimization can not be directly solved using commercial off-the-shelf solvers, thus they are usually converted into a single-level equivalent optimization problem. However, such conversion is difficult since the exact AC OPF models, that appear in the lower level, are nonconvex and thus render many existing techniques inapplicable. The authors of this paper have already explored the single-level reduction approach in a two-part paper \cite{p1} and \cite{p2}, where AC OPF is modeled using a convex quadratically constrained quadratic approximation \cite{cpsota}, but even the most computationally efficient solution techniques start diverging for large systems, i.e. systems with over 70 buses. Here we explore a neural network (NN) metamodeling approach that is known to require significant computational time and resources, but can compute for large systems and even allows for discrete variables in the lower level (at even greater computational cost). On the other hand, the KKT-based single-level reduction techniques solve the upper and the lower level simultaneously so the solution process can diverge for large systems.

Proving the global optimality of solution is not within the scope of this paper. Generally, numerical optimization methods that can find global optimal solutions require much stronger conditions on the optimizing goal function and domain of optimization, for instance, in the case of optimizing a linear function on a convex domain. Our method is general in a sense that it imposes no mathematical conditions on the class of the lower level problem except that, using reasonable time and resources, the lower level can be evaluated a number of times to create a dataset for the NN training.




\subsection{Literature review}

\begin{figure}[!b]
  \centering
  \includegraphics[scale=0.45]{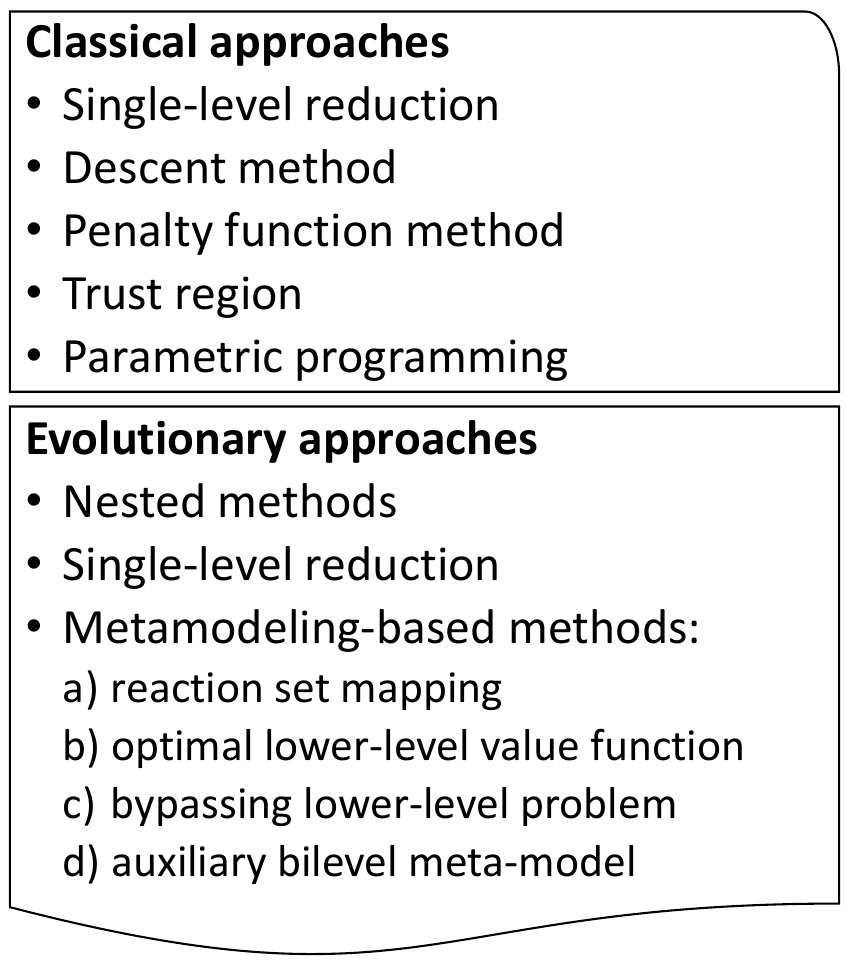}
  \caption{Classification of the bilevel problem solution techniques.}
  \label{fig:literature}
\end{figure}

As depicted in Fig. \ref{fig:literature}, the existing literature on bilevel solution techniques branches out in two main directions: classical and evolutionary approaches. Due to computational difficulty of bilevel problems, the classical approaches can only tackle well-behaved problems with strong assumptions, such as linearity or convex quadraticity and continuity of the lower-level problem, as strong duality generally does not hold for other types of problems. By far the most common classical approach is a single-level reduction based on the KKT conditions and the duality theory. It has been widely used to solve bilevel problems with linear constraints and either linear \cite{KKTlin} or convex quadratic \cite{KKTquad} objective functions, as well as problems with convex quadratic constraints when the interaction between the two levels is discrete \cite{discrete}. The resulting formulations contain complementarity constraints which are combinatorial in their nature and thus can be modeled using binary variables making the final problems mixed-integer linear (MILP) or mixed-integer quadratic (MIQP). The existing state-of-the-art solvers generally handle well these types of optimizations using the branch-and-bound method for binary search tree and simplex for search-tree node subproblems, despite the exponential complexity in the worst case. The single-level reduction technique has also been successfully applied to a case where the lower level is a convex quadratically-constrained quadratic problem (QCQP), as in \cite{BinInteraction}. 
Other classical approaches are the descent method, the penalty function method, the trust-region method and the parametric programming method. The descent method determines the most favorable variable change for the objective function, as demonstrated in \cite{descent}, so the model stays feasible. However, since the model is feasible only when the lower level is optimal, finding the descent direction is very difficult.
The penalty function method replaces the lower-level \cite{SinglePen} or both-level \cite{BothPen} constraints with penalty terms for constraint violations in the objective function. The trust region algorithms iteratively approximate the lower level around the operating point with linear problem (LP) or quadratic problem (QP) \cite{QPTrust}. Both the penalty and the trust-region methods as the next step apply a KKT-based single-level reduction to the lower level and thus inherit the same applicability limits. A recent research thrust in parametric programming has resulted in an alternative approach to solving bilevel programs to global optimality by exploiting the notion of \emph{critical regions}. To this point, solution approaches based on parametric programming have been proposed to handle bilevel programs with LP \cite{param1}, QP \cite{param2}, MILP and MIQP \cite{param3} lower levels.

As opposed to the classical ones, the evolutionary approaches are inspired by the biological evolution principle where candidate solutions are evaluated using a fitness function, e.g. objective function, to form the next generation of candidate solutions by reproducing, mutating, recombining and selecting processes. Evolutionary approaches are very effective at finding good approximate solutions of numerically very difficult problems with fewer regularity assumptions than the classical approaches. For bilevel problems, the evolution is commonly applied in a nested form where the lower level needs to be solved separately for every upper-level solution candidate, as explained and analyzed in \cite{nested}. The upper-level solution candidates are obtained using an evolution, e.g. particle swarm optimization \cite{PSO} or differential evolution \cite{DE}, while the lower level can be solved using classical approaches such as interior point method \cite{hibridDE} or as well using an evolution, as in \cite{DE}. Despite applicability to nonconvex problems, where classical approaches do not hold, the nested evolutionary method does not scale well with the number of upper-level variables as they exponentially increase the number of lower-level optimizations that need to be performed. A single-level reduction technique can also be utilized in the context of evolutionary approaches where numerical evolution concept is applied to the reduced formulation. Due to KKT conditions, this technique inherits regularity assumptions of the classical approaches for the lower-level problem, but allows a more irregular upper level. Paper \cite{EvolutionKKT} is one of the first works where this technique was employed. Evolutionary approaches can also be applied in tandem with the meta-modeling method. Meta-model, or surrogate model, is an easy to evaluate, typically iteratively enhanceable, approximation of the original model. For bilevel modeling, the lower level can be meta-modeled using the reaction set mapping, optimal lower-level value function, by bypassing the lower-level problem completely and by using an auxiliary bilevel meta-model. The reaction set method maps the lower-level variable values as a response to the upper-level variables,as demonstrated in \cite{OptimalSolutionMap}. On the other hand, the optimal lower-level function method replaces the lower objective statement of minimization or maximization with a constraint requiring that the objective is at least as good as the optimal lower-level function \cite{OptValueFun}. Both the reaction set map and the optimal lower-level function are generally difficult to obtain even in an approximated form. Bypassing the lower-level completely is based on the principle that the lower-level variables are basically functions of the upper-level variables, which allows for the upper level reformulation not to include the lower level. Similar to the trust-region method, bilevel problems can be replaced with auxiliary meta-models. As of current, we are not aware of any works based on bypassing the lower-level problem or the auxiliary meta-models methods. A broader bilevel solution techniques research field review, for both the classical and the evolutionary approaches, can be found in \cite{tutorial}.

The approach presented in this paper can be classified as an evolutionary meta-modeling method that bypasses the lower-level problem completely, as given by \cite{tutorial}. This bypassing of the lower-level problem is achieved by approximating the solution of the lower-level, which depends only on the upper-level variables, using a carefully designed NN, see \cite{Knjiga1} and \cite{Knjiga2}. As a NN is simply a function composed of elementary functions, it can be substituted directly into the upper-level objective function. This way, the original bilevel optimization problem is reduced into a single-level optimization problem, which approximates the solution of the original problem. The main difficulty of our framework is the design and training of a NN that efficiently and accurately approximates the lower-level problem.

\subsection{Contribution}

In this work we develop a general numerical solution technique for bilevel problems and apply it to the energy storage (ES) bidding problem on an AC-OPF-constrained energy market. The technique is applicable to any other upper-level subject, but we chose the ES due to modeling simplicity and clarity of presentation. The numerical and mathematical difficulty of solving the considered bilevel optimization arises from insufficient problem regularity due to nonconvexity of the exact AC OPF formulations. Current modeling practice is to avoid the difficulties by using a simpler linear DC OPF \cite{DC} network representation as in \cite{eem}. To the authors knowledge, there are very few attempts to solve bilevel problems with an AC OPF in the lower level. We have not found any with the exact AC OPF, thus we single out two papers with AC OPF relaxations, \cite{relax} and \cite{relax2}. Scalability and tractability issues are not discussed in these papers which is also one of the important points of this work.

The main contribution of this paper consists of the following:
\begin{enumerate}
    \item We introduce a novel numerical scheme for solving bilevel optimization problems based on deep convolutional neural networks. It is an evolutionary meta-modeling method that completely bypasses the lower-level problem. Our method successfully works with previously intractable, i.e. nonconvex, classes of the lower-level problems. As opposed to the existing techniques, solution times are basically independent on the upper-level problem size and scale well with the lower-level problem size.
    \item We demonstrate the solution technique effectiveness by solving a price-maker energy storage AC-OPF-constrained market bidding problem. The results demonstrate higher achieved profits than with the DC market representation.
\end{enumerate}

The paper is organized as follows. Section \ref{sec:math} provides mathematical foundation of the work and is divided in six subsections. Subsection \ref{sub:opt} states the optimization problem, Subsection \ref{sub:LLApp} explains how to approximate the lower level using a neural network, Subsection \ref{sub:fcn} describes the concept of fully connected neural networks, Subsection \ref{sub:conv} explains the advantages, concept and our choice of hyperparameters of the used convolutional neural network, Subsection \ref{sub:train} describes the neural network training algorithm and Subsection \ref{sub:meta-opt} explains our iterative numerical scheme to solve the optimization problem at hand. The case study is presented in Section \ref{sec:case} with implementation details stated in Subsection \ref{sub:case_details} and test results in Subsection \ref{sub:case_tests}. The final Section \ref{sec:conclusion} concludes the paper.

\section{Mathematical Modeling}\label{sec:math}
\subsection{Optimization model}\label{sub:opt}
In the following model we solve optimal ES bidding problem in the AC OPF network-constrained electricity market. The problem is of bilevel structure, i.e. the upper level maximizes the ES profit while the lower level maximizes social welfare due to supply and demand market bids. In the lower level we consider an exact nonconvex quadratic AC OPF formulation based on rectangular coordinates \cite{basic_ACOPF} notation written out in the Appendix, however, other notations such as polar \cite{basic_ACOPF} or current-voltage \cite{IV} are also applicable.

The upper-level problem consists of objective function \eqref{eq:UL.1}, where $\lambdaP$ is the electricity price in each hour indexed by $t$, and $\qbat$ is the average ES power during one hour, i.e. energy, at the interface. Constraint \eqref{eq:UL.2} models the ES (dis)charging process, i.e. change in its state-of-energy $\SoE$ considering charging and discharging efficiencies $\etach$ and $\etadis$. Constraint \eqref{eq:UL.3} sets limits to the state-of-energy (SoE), with $\SoEmax$ being the maximum value. Constraints \eqref{eq:UL.4} and \eqref{eq:UL.5} limit the ES (dis)charged energy to $\qchmax$ for charging and $\qdismax$ for discharging. Binary variable $\xch$ disables simultaneous charging and discharging. Finally, equation \eqref{eq:UL.6} combines charging and discharging into a single variable $\qbat$. Optimization variables are written in formulas in normal font and contained in the variables set $\Xi$, while the parameters are written in bold font.

\begin{equation}\tag{1.1}
\underset{\Xi}{\mathrm{Max}} \: \: - \sum_{t} \qbat \cdot \lambdaP
\label{eq:UL.1}
\end{equation}

\begin{equation}\tag{1.2}
\SoE = \SoEm + \qch\cdot \etach - \qdis/\etadis, \quad \forall{t}
\label{eq:UL.2}
\end{equation}

\begin{equation}\tag{1.3}
0 \le \SoE \le \SoEmax, \quad \forall{t}
\label{eq:UL.3}
\end{equation}

\begin{equation}\tag{1.4}
0 \le \qch \le \qchmax \cdot \xch, \quad \forall{t}
\label{eq:UL.4}
\end{equation}

\begin{equation}\tag{1.5}
0 \le \qdis \le \qdismax \cdot (1-\xch), \quad \forall{t}
\label{eq:UL.5}
\end{equation}

\begin{equation}\tag{1.6}
\qbat = \qch - \qdis, \quad \forall{t}
\label{eq:UL.6}
\end{equation}


The lower level is only textually explained and not written here since we are bypassing it completely. The rectangular AC OPF consists of the objective function, the bus power balance constraints, the power flow equations, the line apparent power limits, the bus voltage limits, the generator production limits and the reference bus constraints. Mathematically challenging are the power flow equations and the lower bus voltage limit constraints, which do not conform to the traditional single-level reduction technique as they are nonconvex. Moreover, there are two additional convex, but nonlinear parts of the formulation. The considered objective function has quadratic cost coefficients and line apparent power limit constraints are of second-order cone form. Broader insights of different AC OPF formulations can be found in tutorial works such as \cite{basic_ACOPF}.

The concept of bypassing the lower level completely is based on the fact that the locational marginal prices $\lambdaP$ are essentially a function of the upper-level $\qbat$ variables, i.e. the objective can be expressed as $\eF\qbatvect$, where $|\tau|$ is the cardinality of set $\tau$ of time steps. However, $\eF$ can not be expressed explicitly, so we replace the objective function  $\eF$ with the approximation $\eFapprox$ from \eqref{eq:UL.7}. The approximating function $\eFapprox$ is given as the feed-forward neural network, so it can be expressed explicitly in terms of elementary mathematical functions. Essentially, we are solving a single-level optimization meta-model that maximizes \eqref{eq:UL.7} subject to constraints \eqref{eq:UL.2}--\eqref{eq:UL.6}.

\begin{equation}\tag{1.7}
\underset{\Xi}{\mathrm{Max}} \: \: \eFapprox\qbatvect
\label{eq:UL.7}
\end{equation}
The problem belongs to the mixed-integer nonlinear optimization class due to the nonlinear neural network function $\eFapprox$ and due to $\xch$ being binary variables.

\subsection{Lower level approximation using neural networks}
\label{sub:LLApp}
It is well known that for any function $f:U\subseteq\mathbb{R}^n\to\mathbb{R}^m$ there exists a NN that uniformly approximates the given function, see \cite{D1} and \cite{D2}. What is typically not known is how exactly to construct a specific NN, approximating the function $f$ to the desired accuracy and using the smallest possible number of neurons.

The first problem we encountered is the limited size of dataset used to train such NN. More precisely, each element in the dataset must be constructed by solving a single instance of the lower-level optimization problem, for some chosen values of the upper-level variables. Solving too many instances of the lower-level problem would take too long. On the other hand, the size of the dataset limits the maximum network size (the number of neurons), by limiting the number of parameters that define that particular network. NNs trained on a dataset that is small compared to the number of network parameters tend to overfit the training data and are poor in generalization on unseen data. In our case that would lead to lower accuracy of approximation of the lower-level problem solutions. Basically, the size of the dataset limits the accuracy of the NN approximation.

The second issue is in determining an optimal topology of a NN for a given network size, in order to achieve the greatest possible approximation accuracy. The optimal topology is dependent on an unknown function, which we are trying to approximate. Our first approach, using fully connected neural network with only few hidden layers, led to poor approximation accuracy. By carefully analyzing the properties of the lower-level optimization problem, the choice of a network topology was settled on a convolutional neural network (CNN) \cite{D3}. As the CNN architecture shares the same values of parameters between different parts of the network, the cumulative number of parameters is much smaller for the network of the same size, so the CNN architecture can be trained to approximate the original optimization problem to a higher accuracy. The first big success of the CNN architecture was in the area of computer vision, in the image classification problems \cite{D4}.

The third obstacle we encountered was the generation of a dataset for the CNN training. Our first idea was to generate the dataset by uniform random sampling of the independent upper-level variables, only in intervals of their permissible values. Then for each sample, we solved the associated lower-level optimization problem. This strategy proved to be inefficient as the near-optimal values of the upper-level variables, which solve our bilevel optimization problem, are poorly represented by sampling these variables independently from the uniform distribution. It resulted in much higher approximation error of the CNN on the optimal solution than on the generated dataset. The solution proved to be in iterative refining of the generated dataset. In the first iteration we generate a uniform dataset on the whole permissible domain and find the solution of the approximation for the bilevel optimization problem. In each additional iteration we restrict the domain to an even smaller neighborhood around the approximated solution from the previous iteration. Then we generate a new uniform dataset on this smaller domain, train a new instance of the CNN, and using this new trained network, we again find a solution of the approximating problem. In each iteration we verify the quality of the current solution by computing the upper-level objective function exactly on optimal variables approximate problem values. We stop iterating when the actual value of the upper-level objective function stops improving.

\subsection{Feed-forward fully connected neural networks}\label{sub:fcn}

\begin{figure}[!b]
  \centering
  \includegraphics[scale=0.7]{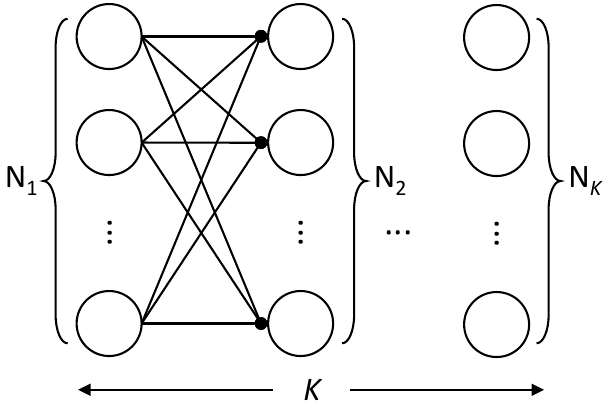}
  \caption{Example of a fully connected neural network.}
  \label{fig:fcnn}
\end{figure}

\begin{figure}[!t]
 \centering
 \includegraphics[scale=0.75,trim={0.88cm 0pt 0.8cm 0pt}]{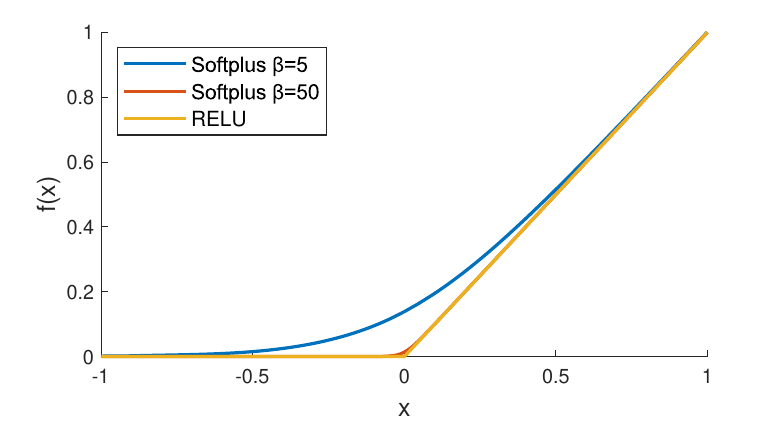}
 \caption{Softplus and RELU plot.}
 \label{fig:softplus}
\end{figure}

\begin{figure*}[!bth]
  \centering
  \includegraphics[scale=0.5]{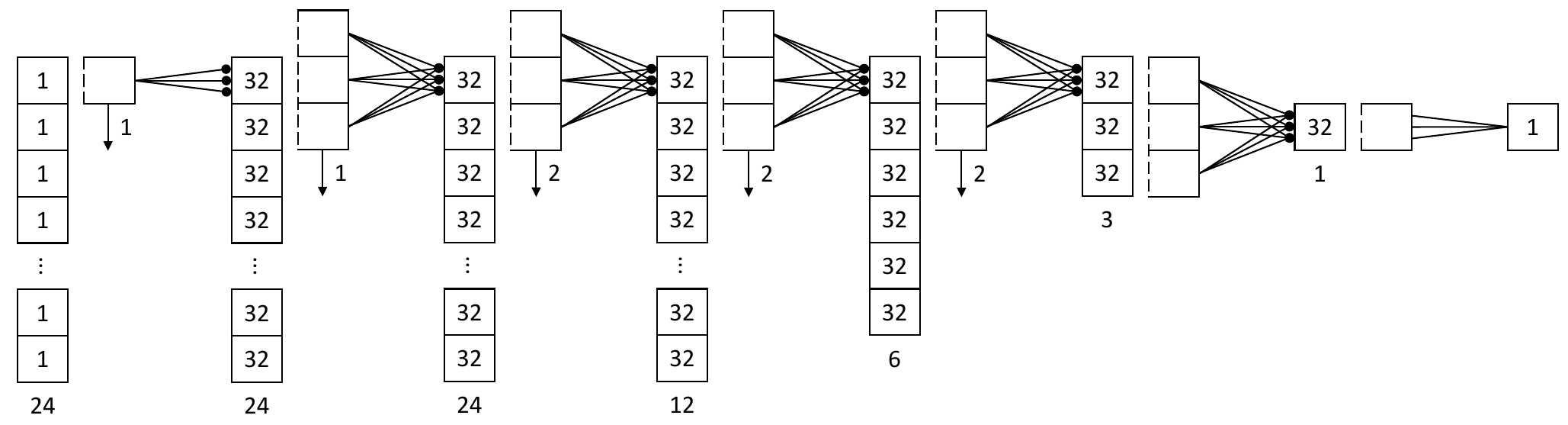}
  \caption{Deep convolutional neural network structure.}
  \label{fig:convolution}
\end{figure*}

A feed-forward fully connected NN (see Figure \ref{fig:fcnn}) consists of $K$ layers, where each layer consists of a number of neurons \cite{D5}. The first layer is referred to as the input layer, the last layer as the output layer, while the intermediate layers are called hidden layers. Neurons in each layer are connected only to the neurons in the neighboring layers. Feed-forward means that the data flows from the input layer to the output layer, strictly from one layer to the next one and in only one direction. Fully connected means that each neuron is connected to every neuron in the neighboring layers. Finally, each neuron in every hidden layer performs a nonlinear transformation on the data by applying the so-called activation function. More precisely, a feed-forward fully connected NN is a function $\mathcal{F}:\mathbb{R}^{N_1}\to\mathbb{R}^{N_K}$, where $N_1$ is the number of neurons in the input layer, and $N_K$ is the number of neurons in the output layer. Function $\mathcal{F}$ is a composition of the alternating affine maps $\mathcal{A}_k:\mathbb{R}^{N_{k-1}}\to\mathbb{R}^{N_k}$, $k=2,\ldots,K$ and the element-wise nonlinear activation functions $\mathcal{N}_k:\mathbb{R}^{N_k}\to\mathbb{R}^{N_k}$, $k=2,\ldots,K-1$, such that
$$
\mathcal{F} = \mathcal{A}_K\circ\mathcal{N}_{K-1}\circ\mathcal{A}_{K-1}\circ\cdots\circ\mathcal{N}_3\circ\mathcal{A}_{3}\circ\mathcal{N}_{2}\circ\mathcal{A}_2 ,
$$
where $N_k$ is the number of neurons in $k$-th layer.

Affine map $\mathcal{A}_k$ can be written in a matrix form as
$$
\mathbf{\hat{z}}_k := \mathcal{A}_k(\mathbf{z}_{k-1}) = \mathbf{W}_k \mathbf{z}_{k-1} + \mathbf{b}_k,\quad\forall k=1,\ldots,K ,
$$
where weight matrix $\mathbf{W}_k$ has dimension $N_k\times N_{k-1}$ and bias vector $\mathbf{b}_k$ has dimension $N_k$.

For the activation functions we element-wise use the Softplus function,
$$
z_k^i:=\mathcal{N}_k^i(\mathbf{\hat{z}}_k) = \frac{\ln(1+\exp(\beta\cdot\hat{z}_k^i)}{\beta},\begin{tabular}{c} $\forall k=1,\ldots,K,$\\ $\forall i=1,\ldots,N_k,$\end{tabular}
$$
where $\hat{z}_k^i$ is the $i$-th component of vector $\mathbf{\hat{z}}_k$ and $\beta$ is the hyperparameter of the Softplus function. Notice that for large values of $\beta$, Softplus uniformly converges to a rectified linear unit (ReLU) activation function (see Figure \ref{fig:softplus}), which is given element-wise by
$$
z_{k}^i=\max(\hat{z}_k^i, 0),\quad\forall k=1,\ldots,K,\quad\forall i=1,\ldots,N_k .
$$
ReLU activation function is commonly used in recent NN applications. The reason why we decided to use Softplus will be become clear in Section \ref{sec:math:meta_opti}.

\subsection{Convolutional neural networks}\label{sub:conv}
\label{section:conv}

A CNN can be regarded as a sub-type of a feed forward NN. It is generally not fully connected, and a large number of weight and bias elements of matrices $\mathbf{W}_k$ and vectors $\mathbf{b}_k$ share the same values, as affine maps $\mathcal{A}_k$ are defined using the operation of matrix convolution \cite{D3}.

Figure \ref{fig:convolution} depicts a deep CNN, describing the exact NN topology used in approximating function $\eFapprox$ of our problem. The structure of the NN is determined by an educated guess of the authors and by experimentation. Besides the input and output layers, we have six additional hidden layers. Unlike in a general NN, each layer in our CNN is described using the layer length $L_k$ and the number of channels $C_k$. The number of neurons in each layer is given by $N_k=L_k\cdot C_k$ and neurons are grouped in $L_k$ groups of size $C_k$. In Figure \ref{fig:convolution}, a single square depicts one group of neurons. The exact number of neurons within each group (the number of channels $C_k$) is written inside each square. The number of groups in each layer is given by a number written just below each layer. For instance, the total number of neurons in the second layer is equal to $24\cdot 32 = 768$.

To define affine map $\mathcal{A}_k$, each layer in a CNN has an additional integer hyperparameter called the kernel size $S_k$. In Figure \ref{fig:convolution}, only the first (input) layer and the last of the hidden layers have kernel sizes $S_1=1$ and $S_7=1$. All the other hidden layers have kernel size equal to $3$. Notice that the kernel size is not applicable to the output layer, as the output layer only collects the output of the last hidden layer and is not applying any further affine maps. All kernels are depicted by a number of empty squares equal to the kernel size $S_k$. A downward arrow indicates that a kernel window is sliding over the layer in steps, performing a computation of the affine map. This means that the convolution operation can be in each step regarded as a smaller affine map $\hat{\mathcal{A}}_k$ that is defined only between the neurons in the groups covered by the kernel window in layer $k-1$ and the neurons in the single output group in layer $k$. In each step of the convolution operation on layer $k-1$ we use the same map $\hat{\mathcal{A}}_k$.

Each convolution layer has two additional integer hyperparameters called a stride and a padding size. The stride is the number of groups by which each kernel window moves in every step of the computing convolution operation. The first and second layers have the stride equal to $1$ and the third to fifth layers have the stride equal to $2$. This is the reason why the lengths $L_k$ of the fourth to sixth layer are decreasing by a factor of $2$. For the sixth and seventh layers, the stride is not applicable as the convolution operation is trivially performed only in the single possible position. The padding controls whether the kernel window can slide over the side of the layer or not. If we let the kernel windows slide over the side of the layer, as for the second to fifth layer, the padding is equal to $1$ and we substitute zeros for the input in the convolution operation in place of the non-existing data. For the first, sixth and seventh layer, the padding is equal to $0$, which means we do not let the kernel window slide over the side of the layer.

In matrix representation $\mathbf{W}_k$ of affine map $\mathcal{A}_k$, lot of matrix components are equal to zero and lot of other non-zero matrix components share the same values. We actually have, for the kernel size equal to $1$ the block diagonal matrix $\mathbf{W}_k$, and for the kernel size equal to $3$ the block tridiagonal matrix $\mathbf{W}_k$, see Figure \ref{fig:matrix}. Every block is of size $C_k\times C_{k-1}$. For the kernel size equal to $1$, every block in the block diagonal matrix $\mathbf{W}_k$ representing affine map $\mathcal{A}_k$ is exactly the same block. For the kernel size equal to $3$, every $3$ vertical blocks in the block tridiagonal matrix representation are exactly the same blocks. The bias vector $\mathbf{b}_k$ also has repeating components. Regardless on the kernel size, components of $\mathbf{b}_k$ repeat every $C_k$ entries, which means each neuron group in a single layer shares the same biases.

Notice that a CNN, for the same number of neurons, typically has much lower number of parameters defining affine maps $\mathcal{A}_k$, than a fully connected NN. For instance, in our case the number of parameters defining map $\mathcal{A}_3$ is $C_2 S_2 C_3 + C_3 = 3104$, whether the number of parameters defining map $\mathcal{A}_3$ in a fully connected neural network with the same number of neurons would be $N_2 N_3 + N_3 = 590592$.

\begin{figure}%
    \centering
    \subfloat[\centering ]{{
\(
\begin{pmatrix}
A_1&0&\cdots&0\\
0&A_2&\cdots&0\\
\vdots&\vdots&\ddots&\vdots\\
0&0&\cdots&A_n\\
\end{pmatrix}
\) 
}}%
    \subfloat[\centering ]{{
\(\!\!\!
\begin{pmatrix}
B_1&C_1& 0 &  \cdots & 0\\
A_2&B_2&C_2& &  \\
\vdots & \ddots & \ddots & \ddots & \vdots \\
& & A_{n-1} &B_{n-1} &C_{n-1} \\
0 & \cdots & 0 &A_{n} &B_{n} \\
\end{pmatrix}
\) 
}}%
    \caption{Block matrices: a) diagonal; b) tridiagonal}
    \label{fig:matrix}%
\end{figure}

\subsection{Training feed-forward neural networks}\label{sub:train}

To train a neural network simply means to optimize matrices $\mathbf{W}_k$ and vectors $\mathbf{b}_k$ in order to minimize a chosen loss function over a dataset. Since we use CNNs, we must respect the block diagonal or tridiagonal structure of matrices $\mathbf{W}_k$ having many shared values, as discussed in Section \ref{section:conv}.

The dataset is generated by solving a number of instances of the lower-level optimization problem, for as many random samples of vector $\qbatvect$, and subsequently evaluating value of the target objective function \eqref{eq:UL.1}, i.e. $\eF\qbatvect$.

The loss function is the mean squared error between the NN computed value $\eFapprox\qbatvect$ and the lower-level exact solution $\eF\qbatvect$.

As customary in NN training, the optimization of $\mathbf{W}_k$ and $\mathbf{b}_k$ is performed using a variant of a gradient descent optimizer. The dataset is split into training and validation datasets, using $80:20$ percent split ratio. The optimization is done in multiple epochs over the training dataset, where gradients are computed using backpropagation \cite{Nature} and automatic differentiation \cite{AD} of NN. The validation dataset is used only for evaluating the loss function after every epoch of training and not for a gradient computation. Computed values of the loss function on the validation dataset are used to asses numerical viability of the training process and to select the best trained network, having the lowest value of the validation loss. Before the start of the first epoch of training, $\mathbf{W}_k$ and $\mathbf{b}_k$ are initialized to random values.

More details about our exact training procedure, together with all optimizer and training hyperparameter values are given in Section \ref{section:case_study:imp_details}.

\subsection{Meta-optimization numerical scheme}\label{sub:meta-opt}
\label{sec:math:meta_opti}

We devised an iterative numerical scheme for computing a sequence of CNN approximations $\eFapprox_i$ of the otherwise intractable objective function $\eF$. In each iteration $i$ we first generate dataset $\mathcal{D}_i$ of random sampled vectors $\qbativect$. For each $t\in\tau$, values of $\qbati$ are independently and uniformly random sampled from the interval centered around $\qbatcenter$ of length at most $2\qbatrad$, respecting the condition $-\qdis\leq\qbati\leq\qch$. In the first iteration we set $\qbatcenter$ equal to zero and $\qbatrad$ such that it allows all permissible values of $\qbat$, i.e. $-\qdis\leq\qbati\leq\qch$.

In practice, we noticed that our numerical scheme runs better if we introduce an additional small relative tolerance $\epsilon>0$ on the dataset creation. Thus, for the maximum length of the sampling interval we actually use $2\qbatrad(1+\epsilon)$ and allow all permissible values of $\qbat$ to be from the interval $-\qdis(1+\epsilon)\leq\qbati\leq\qch(1+\epsilon)$. The intuition behind introducing the tolerance is that our CNN would better approximate objective function $\eF$ for parameter values $\qbati$ near the boundary values $-\qdis$ and $\qch$, if the dataset is allowed to include values $\qbati$ a bit outside the interval $[-\qdis, \qch]$.

Each element in dataset $\mathcal{D}_i$ is now a pair of random vector $\qbativect$ and a value $F\qbativect$ of the target objective function \eqref{eq:UL.1}, computed by solving a single instance of the lower-level problem.

Next, in each iteration we separately train a number of CNNs, $\eFapprox_{i,n}$ indexed by $n\in\{1, \ldots, M\}$, to approximate function $\eF$. For every additional training on the same dataset $\mathcal{D}_i$ we obtain a subtly different CNN, as the training process is intrinsically stochastic (random initialized network weights and random sampled stochastic gradient descent mini batches).

Now, as every trained CNN is a function $\eFapprox_{i,n}$, we symbolically insert function $\eFapprox_{i,n}$ into the upper-level problem objective function. For this we use our modeling environment's, which is AMPL, defining variables feature. Defining variables are a type of variables that are substituted out, potentially in a nested way, by their declaration expression before reaching the solver. This results in solving a single-level optimization meta-model which maximizes

\begin{equation}\tag{1.8}
\underset{\Xi}{\mathrm{Max}} \: \: \eFapprox_{i,n}\qbativect
\label{eq:UL.8}
\end{equation}
subject to constraints \eqref{eq:UL.2}--\eqref{eq:UL.3} and

\begin{equation}\tag{1.9}
0 \le \qchi \le \qchmax \cdot \xchi, \quad \forall{t}
\label{eq:UL.9}
\end{equation}

\begin{equation}\tag{1.10}
0 \le \qdisi \le \qdismax \cdot (1-\xchi), \quad \forall{t}
\label{eq:UL.10}
\end{equation}

\begin{equation}\tag{1.11}
\qbati = \qchi - \qdisi, \quad \forall{t}
\label{eq:UL.11}
\end{equation}

\begin{equation}\tag{1.12}
\qbatcenter - \qbatrad \le \qbati \le \qbatcenter + \qbatrad, \quad \forall{t} ,
\label{eq:UL.12}
\end{equation}
for every trained CNN indexed by $n$. The additional constraint \eqref{eq:UL.12} is used to respect that dataset $\mathcal{D}_i$ is created centered around $\qbatcenter$ using an interval length at most $2\qbatrad$.

Notice, as we used the differentiable Softplus activation function in our CNN, and as a CNN is just a composition of affine maps and element-wise activation functions, that $\eFapprox_{i,n}$ are differentiable functions. In case of using a ReLU activation function, the resulting $\eFapprox_{i,n}$ would not be differentiable.
We experimentally established that lower values of the Softplus hyperparameter $\beta$ result in lower overall neural network approximation accuracy, and higher values give rise to solver instabilities in the single-level meta-model optimization step. We also tried to use ReLU instead of Softplus, but we were plagued with solver instabilities and slowdown. Using Softplus showed to be much more efficient.

For every trained CNN we now have the computed profit $\compprofitin = \underset{\Xi}{\mathrm{Max}} \: \: \eFapprox_{i,n}(\qbati)$ and the computed optimal ES (dis)charged energy $\qbatinvect$. Now we can verify what are the actual profits obtained for the computed optimal ES (dis)charging schedule. This is done by optimizing the lower-level problem independently of the upper level with fixed ES (dis)charging schedule. The actual profit is determined as $\veriprofitin = \underset{\Xi}{\mathrm{Max}} \: \:  -\sum_{t} \qbatin \cdot \lambdaP$, where $\lambdaP$ is in this case the bus balance constraint marginal, computed by default by many interior point solvers.


Additionally, we compute mean optimal ES energy exchange quantities
\[
\qbatimean = \frac{1}{M}\sum\limits_{n=1}^M \qbatin\quad \forall t\in\tau
\]
and for the computed mean vector $\qbatimeanvect$ we again optimize the lower-level problem with fixed ES charging values to the obtained mean optimal vector, arriving at the mean actual profit $\veriprofitimean = -\sum_{t} \qbatimean \!\cdot\! \lambdaP$. Considering the mean actual profit is justified, because averaging over optimal solutions of many different CNN models $\eFapprox_{i,n}$, each one approximating an intractable objective function $\eF$, it can result in a better mean solution. By looking at the actual test results (i.a.\ Tables \ref{table:profit:3lmbd} and \ref{table:profit:57ieee}) we see that this approach is justified in practice, i.e., in some iterations the mean actual profit $\veriprofitimean$ can be higher than any of the actual profits $\veriprofitin$.

Finally, we have to choose values of $\qbatcenternext$ and $\qbatradnext$ for a next iteration of the meta-optimization scheme. For $\qbatcenternext$ we either chose the optimal ES (dis)charging quantities $\qbatinvect$ from CNN that achieved the highest actual profit $\veriprofitin$, or in the case $\veriprofitimean$ is the highest profit, we chose the mean optimal ES (dis)charging quantities $\qbatimeanvect$.

To choose $\qbatradnext$, we first compute the maximum over all $t\in\tau$ of the standard deviations of samples $\left\{\qbatin : 1\leq n\leq M\right\}$. More precisely, we compute
\[
\sigma_i := \underset{t\in\tau}{\mathrm{Max}}\ \mathrm{std}\left\{\qbatin : 1\leq n\leq M\right\} .
\]
For the next iteration we take a smaller value between the current $\qbatrad$ and a new estimate,
\[
\qbatradnext := \mathrm{Min}\{\qbatrad,\,\gamma\cdot\sigma_i\} ,
\]
for every $t\in\tau$, where $\gamma$ is a hyperparameter. Notice that $\qbatradnext$ does not depend on $t$. It has the same value for every $t$.

For the next iteration of our meta-optimization scheme, dataset $D_{i+1}$ is created around $\qbatcenternext$, which is the best optimal ES (dis)charging schedule computed in the current iteration, i.e., ES charging and discharging energy quantities that produce the highest profit. The dataset width, which is decided by $\qbatradnext$, is influenced by how close together are optimal ES schedules predicted by different CNNs trained in the current iteration. In case different CNNs produce relatively close optimums, the computed standard deviation $\sigma_i$ is relatively small, and a next iteration dataset is going to be concentrated around a smaller neighborhood of $\qbatcenternext$. On the contrary, in case different CNNs produce optimums that are more apart, the computed standard deviation $\sigma_i$ is relatively large, and a next iteration dataset is going to span over a bigger neighborhood of $\qbatcenternext$. Notice that $\qbatrad$ is non-increasing between iterations.

In the end, we have to prescribe a stopping criterion for our meta-optimization scheme. We choose to stop further iterations if there is no improvement in the actual profit of the current iteration compared to the previous one. We empirically conclude (see Section \ref{sub:case_tests}) that the convergence of our meta-optimization scheme is achieved in few iterations (see Tables \ref{table:profit:3lmbd}, \ref{table:profit:57ieee} and \ref{table:profit:73ieee}). An overview of the numerical optimization scheme is provided in Algorithm \ref{alg:scheme}.

\begin{algorithm}[bht]
  \caption{Numerical optimization scheme}\label{alg:scheme}
  \begin{algorithmic}[1]
    \Repeat
      \State Generate a new random dataset ($10^5$ entries)
      \State Evaluate LL response for the dataset
      \State Train 60 NNs to approximate LL response
      \State Optimize the ULs with inserted NNs into objective
      
      \hspace{-0.3cm} function
      \State Determine actual profits by optimizing LL with fixed
      
      \hspace{-0.3cm} ES (dis)charging schedule
      \State Select the best actual solution out of:
      \begin{itemize}[leftmargin=3em]
          \item the best direct result;
          \item the result obtained averaging decisions from all optimized NNs;
      \end{itemize}
      \State For the next iteration, reduce and concentrate the
      
      \hspace{-0.3cm} dataset spatial size in the neighborhood of the best
      
      \hspace{-0.3cm} solution found from this iteration
      \Until{The best solution is worse than in the preceding iteration}
  \end{algorithmic}
\end{algorithm}

\section{Case Study}\label{sec:case}

\subsection{Implementation details}\label{sub:case_details}
\label{section:case_study:imp_details}

For the dataset creation, each instance of the lower-level problem, one for every dataset entry, was solved using AMPL running KNITRO 12.3 solver. A single dataset entry consists of a 24-dimensional floating point vector $\qbatvect$ as an input and a single floating point value as an output, which corresponds to the computed upper-level profit for given $\qbatvect$ values (computed as $-\sum_{t} \qbat \cdot \lambdaP$, where $\lambdaP$ is the marginal of an active power bus balance constraint at the ES location). In total, we used $10^5$ dataset entries and thus the same number of independent lower levels to be solved. To reduce the computation time, computations were carried out in parallel running on a dual Intel Xeon CPU computer system over a total of 40 physical cores.

We implemented our CNN depicted in Figure \ref{fig:convolution} in Python using PyTorch library \cite{D6}. The exact hyperparameters of our CNN are already described in detail in Section \ref{section:conv}. Regarding the remaining hyperparameter $\beta$ of the Softplus activation function, after conducting extensive training experiments, we settled on the value $\beta=50$.

Training of the CNN was implemented using fast.ai library \cite{fastai}, which allowed us to easily implement several state-of-the-art optimizer features, likely improving over the nowadays rather standard Adam \cite{D7} and stochastic gradient descent optimizers.

Overall, the training procedure was very similar to the one employed in \cite{DD}. For the optimizer we used the Ranger algorithm \cite{Ranger}, which can stabilize the start of the training by using the Rectified Adam (RAdam) optimizer \cite{D8}. Another feature we used is the parameter lookahead, which avoids overshooting a good local minima in parameter space \cite{D9}, thus stabilizing the rest of the training process. The learning rate and weight decay were controlled during the training using the flat-cosine one-cycle policy. This produces convergence to a broader optimum, allowing better generalization for the trained neural network \cite{D10}.

RAdam has several important hyperparameters that greatly influence the speed of the training convergence and the quality of the obtained trained neural network in respect to generalizability. These hyperparameters are: the number of epochs of training, the maximum learning rate, the training batch size, the weight decay factor and the exponential decay rates of the first and second moments. We experimented with training thousands of models in order to find the combination of the optimizer hyperparameters that work well for our CNN topology and dataset properties. Finally, we fixed the number of epochs to $500$, maximum learning rate to $0.003$, training batch size to $128$, weight decay factor to $0.01$ and exponential decay rates of the first and second moments to values $0.95$ and $0.85$.

In every iteration of the meta-optimization scheme we trained a total of $M=60$ CNNs on the same dataset. Our computer system was equipped with 6 Nvidia Quadro RTX 6000 GPUs, each having 16 Gb of RAM. As the CNN and dataset sizes were relatively small and used only a fraction of our GPU memory and computational power, we could efficiently train all CNNs in parallel, training 10 CNNs per GPU. The slowdown induced by parallel training of 10 models per GPU is insignificant compared to alternatively much longer time of sequential single-model training.

In the meta-optimization scheme there are two hyperparameters to consider. After experimenting with different values, for the relative tolerance of the dataset creation we take $\epsilon=0.1$, and for the other hyperparameter we take $\gamma=5$. A value of hyperparameter $\gamma$ influences the decreasing rate of $\qbatrad$ through subsequent iterations. Using lower values of $\gamma$ produces lower values of $\qbatrad$ in later iterations, which can lead to a sub-optimal optimization result in the end. The chosen value of $\gamma$ is a compromise, in a sense that a higher value tends to slow down the speed of convergence. By using $\gamma=5$, the reduction of $\qbatrad$ is not overly aggressive. On the other end, it stabilizes our meta-optimization scheme, which leads to finding better optimums. Table \ref{table:qbatrad} shows the $\qbatrad$ decrease throughout all iterations from the case study. From the table we see that the radius decrease is significant, but not too sudden.

\begin{table}[b]
\caption{Values of $\qbatrad$ for different transmission networks and iterations}
\label{table:qbatrad}
\centering
\small
\begin{tabular}{|c|c|ccc}
\hline
Iter & 3\_lmbd & \multicolumn{1}{c|}{57\_ieee} & \multicolumn{1}{c|}{73\_ieee\_rts} & \multicolumn{1}{c|}{300\_ieee} \\ \hline
1    & 0.6     & \multicolumn{1}{c|}{0.6}      & \multicolumn{1}{c|}{0.6} & \multicolumn{1}{c|}{0.6}           \\ \hline
2    & 0.0618  & \multicolumn{1}{c|}{0.3265}   & \multicolumn{1}{c|}{0.4894} & \multicolumn{1}{c|}{0.2302}        \\ \hline
3    & 0.0319  & \multicolumn{1}{c|}{0.2123}   & \multicolumn{1}{c|}{0.3465}        \\ \cline{1-4}
4    & 0.0129  & \multicolumn{1}{c|}{0.1876}   & \multicolumn{1}{c|}{0.3465}        \\ \cline{1-4}
5    & 0.0129  &                               &                                    \\ \cline{1-2}
6    & 0.0129  &                               &                                    \\ \cline{1-2}
\end{tabular}
\end{table}


\begin{table}[b]
\caption{Average processing time in seconds per single iteration of the meta-optimization scheme.}
\label{table:time}
\centering
\small
\setlength{\tabcolsep}{2pt} 
\begin{tabular}{l|c|c|c|c|}
\cline{2-5}
                                    & \begin{tabular}[c]{@{}c@{}}Dataset\\ creation\\ {[}s{]}\end{tabular} & \begin{tabular}[c]{@{}c@{}}NN training\\ {[}s{]}\end{tabular} & \begin{tabular}[c]{@{}c@{}}Solving\\ meta-models\\ (mean $\pm$ std) {[}s{]}\end{tabular} & \begin{tabular}[c]{@{}c@{}}Component \\ total time\\ {[}s{]}\end{tabular} \\ \hline
\multicolumn{1}{|l|}{3\_lmbd}       & 471                                                                  & 5118                                                          & 60$\times (7.7\pm 10.0)$                                                                 & 6051                                                    \\ \hline
\multicolumn{1}{|l|}{57\_ieee}      & 1774                                                                 & 5129                                                          & 60$\times (11.7\pm 10.6)$                                                                & 7605                                                    \\ \hline
\multicolumn{1}{|l|}{73\_ieee\_rts} & 3644                                                                 & 5125                                                          & 60$\times (4.0\pm 2.6)$                                                                  & 9009                                                    \\ \hline
\multicolumn{1}{|l|}{300\_ieee} & 28256                                                                 & 5245                                                          & 60$\times (2.0\pm 1.9)$                                                                  & 33621                                                    \\ \hline
\end{tabular}
\end{table}

\subsection{Test results}
\label{sub:case_tests}

We tested our method on three separate transmission system meshed networks from PGLib-OPF \cite{pglib} library: 3\_lmbd, 57\_ieee and 73\_ieee\_rts. A time dimension was added to the data by applying the load scaling factors for winter workdays available from IEEE RTS-96 \cite{rts96}. Set of time steps $\tau$ has $24$ elements for different hours in a single day. In case of 73\_ieee\_rts network, we also applied 0.85 scaling factor to the transmission lines capacities to induce congestion. The networks were otherwise unmodified. For an ES to have an impact on the energy market prices, a feature for which the bilevel modeling is used for, it has to be very large. Thus, we model the ES with 100 MWh (1 p.u.) capacity. Charging and discharging efficiencies were both set to 90\% and maximum ES (dis)charging power to 60 MW.

In Table \ref{table:time} we present an average wall time per iteration and for different steps of our meta-optimization scheme. Dataset creation is a cumulative time for three sub-steps: generation of $10^5$ random vectors of $\qbat$, solving lower-level problems for every dataset entry, and data format post-processing of the generated dataset, which mostly include disk input-output (IO) operations. Most of the time is consumed for solving $10^5$ lower-level problems. Notice that bigger transmission system networks require more solver time. NN training is the time consumed for parallel CNN models training. Notice that this time does not depend on the transmission system network size, as it depends solely on the CNN and training hyperparameters. The last step is solving the meta-models, which is done sequentially for all 60 trained CNNs. A possible speedup of using parallel computations in this step would not be significant compared to the total time used per iteration. Notice that the time for solving the meta-models can vary greatly between different CNNs and different iterations. We consider this to be a normal solver behavior occurring due to binary variables $\xch$. For the dataset creation and the NN training, the average wall time is pretty much unchanged between different iterations, so we supply only the mean times without the standard deviation. Our model is highly scalable as long as the lower-level problem can be evaluated a number of times under reasonable time and resources. The alternative method from our two-part paper \cite{p1} and \cite{p2} has scalability issues when using lower levels with larger networks since it computes the lower level and the upper level simultaneously so the solution process can diverge. As seen in Table II, our method from this paper scales reasonably even for 73 and 300 bus systems.

\begin{table}[b]
\setlength{\tabcolsep}{1.5pt} 
\caption{Profits for 3\_lmbd network (ES at bus 3).}
\label{table:profit:3lmbd}
\centering
\small
\begin{tabular}{|c|c|c|c|c|c|c|}
\hline
Iter & \begin{tabular}[c]{@{}c@{}}Total \\ time \\ {[s]}\end{tabular} &  \begin{tabular}[c]{@{}c@{}}Mean $\qbat$\\ actual\\ profit\end{tabular} & \begin{tabular}[c]{@{}c@{}}Best NN\\ actual\\ profit\end{tabular} & \begin{tabular}[c]{@{}c@{}}Best NN\\ computed\\ profit\end{tabular} & \begin{tabular}[c]{@{}c@{}}Single-level\\ reduction\\ actual profit\\ \cite{p1}, \cite{p2} \end{tabular} & \begin{tabular}[c]{@{}c@{}}DC OPF\\ actual\\ profit\end{tabular} \\ \hline
1    & 8386 &  2016.0583                                                              & \textbf{2016.2954}                                                & 2011.5968                                                          & \multirow{6}{*}{2016.8762} & \multirow{6}{*}{1986.4979}                                       \\ \cline{1-5}
2    & 7158 &  \textbf{2016.8395}                                                     & 2016.7695                                                         & 2018.1744                                                        &   &                                                                  \\ \cline{1-5}
3    & 6965 &  \textbf{2016.8414}                                                     & 2016.8271                                                         & 2017.5707                                                        &   &                                                                  \\ \cline{1-5}
4    & 6910 &  \textbf{2016.8416}                                                     & 2016.8339                                                         & 2016.9533                                                        &   &                                                                  \\ \cline{1-5}
5    & 6903 &  \textbf{2016.8505}                                                     & 2016.8015                                                         & 2017.2495                                                        &   &                                                                  \\ \cline{1-5}
6    & 6882 &  2016.8452                                                     & 2016.8096                                                         & 2017.1350                                                           &     &                                                             \\ \hline
\end{tabular}
\end{table}

In Tables \ref{table:profit:3lmbd}, \ref{table:profit:57ieee}, \ref{table:profit:73ieee} and \ref{table:profit:300ieee} we present optimization results in terms of the ES computed and actual profits acquired in four different transmission systems. Actual profits are obtained in the verification Step 6 of Algorithm \ref{alg:scheme} by optimizing the lower level with fixed ES charging decisions as explained in Section \ref{sec:math:meta_opti}. We also compare actual profits achieved by our method to actual profits achieved by solving a bilevel ES market optimal bidding using the AC OPF model and single-level reduction approach from the two-part paper \cite{p1} and \cite{p2} and using a standard DC OPF \cite{DC} model in the lower level. AC OPF single-level reduction approach results in slightly higher actual profits compared to NN approach, but its solution process fails to converge for 73- and 300-bus networks. Actual DC OPF profits are profits that would occur in the AC OPF market, but by using bidding decisions from the DC OPF bilevel model. In our tables, the best NN computed profit is the maximum value of $\compprofitin$ over all $60$ NNs, the best NN actual profit is the maximum value of $\veriprofitin$ over all $60$ NNs, and the mean $\qbat$ actual profit is $\veriprofitimean$, where $i$ is the iteration number and the NN $n$. By design, the best profit is always achieved in the penultimate iteration of our method, as a worse profit in the last iteration actually triggers the stopping criterion. The number of required iterations differs between the transmission systems and ranges from $2$ to $6$ iterations. Profit increase over the DC OPF model was $1.5\%$ for 3\_lmbd, $9.3\%$ for 57\_ieee, $11.8\%$ for 73\_ieee network, and $16.4\%$ for 300\_ieee network. From these tables it is observable that we also achieved a high first iteration accuracy, since the greatest second iteration improvement of the actual profit is only 0.03\%. Notice that the total time per iteration presented is somewhat higher than a component total time in Table \ref{table:time}, as it includes an additional overhead for some data reformatting and IO disk operations. Also, notice that we decided to present profits using up to four decimal places, so that small improvements between subsequent iterations in Table \ref{table:profit:3lmbd} become visible, which also reaffirms the optimality of the first iteration result.

\begin{table}[!t]
\caption{Profits for 57\_ieee network (ES at bus 1).}
\label{table:profit:57ieee}
\centering
\small
\setlength{\tabcolsep}{1.5pt} 
\begin{tabular}{|c|c|c|c|c|c|c|}
\hline
Iter & \begin{tabular}[c]{@{}c@{}}Total \\ time \\ {[s]}\end{tabular}  & \begin{tabular}[c]{@{}c@{}}Mean $\qbat$\\ actual\\ profit\end{tabular} & \begin{tabular}[c]{@{}c@{}}Best NN\\ actual\\ profit\end{tabular} & \begin{tabular}[c]{@{}c@{}}Best NN\\ computed\\ profit\end{tabular} & \begin{tabular}[c]{@{}c@{}}Single-level\\ reduction\\ actual profit \\ \cite{p1}, \cite{p2}\end{tabular} & \begin{tabular}[c]{@{}c@{}}DC OPF\\ actual\\ profit\end{tabular} \\ \hline
1    & 8305  & 1564.0254                                                              & \textbf{1564.1351}                                                & 1564.3315                                                      &\multirow{4}{*}{1565.2053}     & \multirow{4}{*}{1431.5387}                                       \\ \cline{1-5}
2    & 8724  & 1564.5929                                                              & \textbf{1564.6218}                                                & 1572.8308                                                           & &                                                                 \\ \cline{1-5}
3    & 8886 & \textbf{1564.9676}                                                     & 1564.8713                                                         & 1568.1618                                                        &   &                                                                  \\ \cline{1-5}
4    & 9088 & 1564.9433                                                              & 1564.9370                                                         & 1570.1936                                                       &    &                                                                  \\ \hline
\end{tabular}
\end{table}

\begin{table}[!t]
\caption{Profits for 73\_ieee\_rts network (ES at bus 101).}
\label{table:profit:73ieee}
\centering
\small
\setlength{\tabcolsep}{1.2pt} 
\begin{tabular}{|c|c|c|c|c|c|c|}
\hline
Iter & \begin{tabular}[c]{@{}c@{}}Total \\ time \\ {[s]}\end{tabular} & \begin{tabular}[c]{@{}c@{}}Mean $\qbat$\\ actual\\ profit\end{tabular} & \begin{tabular}[c]{@{}c@{}}Best NN\\ actual\\ profit\end{tabular} & \begin{tabular}[c]{@{}c@{}}Best NN\\ computed\\ profit\end{tabular} & \begin{tabular}[c]{@{}c@{}}Single-level\\ reduction\\ actual profit \\ \cite{p1}, \cite{p2}\end{tabular} & \begin{tabular}[c]{@{}c@{}}DC OPF\\ actual\\ profit\end{tabular} \\ \hline
1    & 10234  & 5503.2526                                                              & \textbf{5512.4143}                                                & 5525.5078                                                      & \multirow{4}{*}{No solution}     & \multirow{4}{*}{4927.1127}                                       \\ \cline{1-5}
2    & 10186 &  5503.2608                                                              & \textbf{5512.6547}                                                & 5520.6102                                                        &   &                                                                  \\ \cline{1-5}
3    & 10159 & 5503.8406                                                              & \textbf{5512.7558}                                                & 5519.5750                                                       &    &                                                                  \\ \cline{1-5}
4    & 10192   & 5503.7870                                                              & 5512.1032                                                         & 5513.6247                                                     &      &                                                                  \\ \hline
\end{tabular}
\end{table}

\begin{table}[!t]
\caption{Profits for 300\_ieee network (ES at bus 1).}
\label{table:profit:300ieee}
\centering
\small
\setlength{\tabcolsep}{1.2pt} 
\begin{tabular}{|c|c|c|c|c|c|c|}
\hline
Iter & \begin{tabular}[c]{@{}c@{}}Total \\ time \\ {[s]}\end{tabular} & \begin{tabular}[c]{@{}c@{}}Mean $\qbat$\\ actual\\ profit\end{tabular} & \begin{tabular}[c]{@{}c@{}}Best NN\\ actual\\ profit\end{tabular} & \begin{tabular}[c]{@{}c@{}}Best NN\\ computed\\ profit\end{tabular} & \begin{tabular}[c]{@{}c@{}}Single-level\\ reduction\\ actual profit \\ \cite{p1}, \cite{p2}\end{tabular} & \begin{tabular}[c]{@{}c@{}}DC OPF\\ actual\\ profit\end{tabular} \\ \hline
1    & 33482  & 1396.5797                                                              & \textbf{1397.0175}                                                & 1377.6627                                                      & \multirow{2}{*}{No solution}     & \multirow{2}{*}{1199.9745}                                       \\ \cline{1-5}
2    & 36481 & 1396.7220                                                              & 1397.0172                                                & 1399.9574                                                        &   &                                                                  \\ \hline
\end{tabular}
\end{table}

\section{Conclusion}\label{sec:conclusion}


In this paper we present a novel numerical method, which utilizes deep convolutional neural networks to efficiently solve a wide class of bilevel optimization problems arising in deregulated power systems. Our method uses evolutionary meta-modeling to bypass the lower-level problem completely, thus it is insensitive to the lower-level complexity, which is the main culprit in rendering bilevel optimization problems intractable. We can successfully deal with nonlinear nonconvex lower-levels that include binary variables, as long as the lower-level can be efficiently solved as a single-level problem by treating all upper level variables as parameters. We used our method to specifically solve ES market participation problem using an AC OPF model in the lower level, which would enable the market operators to perform market clearing using the AC OPF instead of much less accurate DC OPF. However, the proposed framework is generally applicable for any other bilevel optimization problem in the power systems domain.

Additionally, our method is scalable in terms of the required precision versus the run time. Using larger training datasets we could train a neural network having more parameters than we used here. We would obtain more accurate optimums, but would also require longer run time. On the other hand, if we are satisfied with lower precision, we could use a smaller dataset for training a smaller neural network, and our method would run faster.

Finally, we note that this procedure can be easily implemented to trilevel models, as they are generally solved by first merging the middle- and the lower-level problems into a mixed-integer problem with equilibrium constraints, see e.g. \cite{naja}, which is a direct application of our proposed procedure. The obtained single-level problem then acts as a lower-level problem to the original upper-level problem. The resulting bilevel structure is commonly iteratively solved using a cutting plane algorithm. However, a direct implementation of our procedure to trilevel problems will be explored in our future work.


\newcommand{\AF}{E}
\newcommand{\AT}{E^\mathrm{R}}
\newcommand{\A}{\AF \!\cup \!\AT}
\newcommand{\BP}{N^{\mathrm{P}}}
\newcommand{\BL}{L_i}
\newcommand{\BS}{S_i}
\newcommand{\G}{G}
\newcommand{\BG}{G_i}
\newcommand{\RB}{R}

\newcommand{\BPR}{N^{\mathrm{PR}}}

\newcommand{\BB}{\beta}

\renewcommand{\wp}{op}
\newcommand{\es}{es}

\newcommand{\CC}{\bm{\ddot{c}}_{k}}
\newcommand{\C}{\bm{\dot{c}}_{k}}
\newcommand{\cc}{\bm{c}_{k}}
\newcommand{\Pgen}{P^{\mathrm{g}}_{t,k}}
\newcommand{\Qgen}{Q^{\mathrm{g}}_{t,k}}
\newcommand{\Pgenm}{P^{\mathrm{g}}_{t-1,k}}
\newcommand{\Qgenm}{Q^{\mathrm{g}}_{t-1,k}}
\renewcommand{\P}{P_{t,e,i,j}}
\newcommand{\Q}{Q_{t,e,i,j}}
\newcommand{\Sline}{\bm{\overline{S}}_{e}}
\newcommand{\dP}{\bm{P}^{\mathbf{d}}_{t,l}}
\newcommand{\dQ}{\bm{Q}^{\mathbf{d}}_{t,l}}
\newcommand{\Vtn}{\bm{V}^{\mathbf{\wp}}_{t,i}}
\newcommand{\Vtm}{\bm{V}^{\mathbf{\wp}}_{t,j}}
\newcommand{\fitn}{\bm{\theta}^{\mathbf{\wp}}_{t,i}}
\newcommand{\fitm}{\bm{\theta}^{\mathbf{\wp}}_{t,j}}
\newcommand{\dVn}{V^{\Delta}_{t,i}}
\newcommand{\dVm}{V^{\Delta}_{t,j}}
\newcommand{\dfin}{\theta^{\Delta}_{t,i}}
\newcommand{\dfim}{\theta^{\Delta}_{t,j}}

\newcommand{\Va}{\widecheck{V}_{t,e}}

\newcommand{\cosf}{\widehat{cos}_{t,i,j}}
\newcommand{\cost}{\widehat{cos}_{t,j,i}}

\newcommand{\gl}{\bm{g}_{e}}
\newcommand{\gf}{\bm{g}^{\mathbf{fr}}_{e}}
\newcommand{\gt}{\bm{g}^{\mathbf{to}}_{e}}

\newcommand{\bl}{\bm{b}_{e}}
\renewcommand{\bf}{\bm{b}^{\mathbf{fr}}_{e}}
\newcommand{\bt}{\bm{b}^{\mathbf{to}}_{e}}

\newcommand{\gs}{\bm{g}^{\mathbf{sh}}_{s}}
\newcommand{\bs}{\bm{b}^{\mathbf{sh}}_{s}}

\newcommand{\tap}{\bm{\tau}_{e}}
\newcommand{\tr}{\bm{\tau}_{e}^{\mathbf{r}}}
\newcommand{\ti}{\bm{\tau}_{e}^{\mathbf{i}}}

\newcommand{\shift}{\bm{\sigma}_{e}}

\newcommand{\Pmax}{\overline{\bm{P}}^{\mathbf{g}}_k}
\newcommand{\Pmin}{\underline{\bm{P}}^{\mathbf{g}}_k}
\newcommand{\Qmax}{\overline{\bm{Q}}^{\mathbf{g}}_k}
\newcommand{\Qmin}{\underline{\bm{Q}}^{\mathbf{g}}_k}
\newcommand{\Vmax}{\overline{\bm{V}}_{i}}
\newcommand{\Vmin}{\underline{\bm{V}}_{i}}

\newcommand{\cospsinij}{\bm{cps}_{t,e,i,j}}
\newcommand{\cospsinji}{\bm{cps}_{t,e,j,i}}
\newcommand{\cosmsinij}{\bm{cms}_{t,e,i,j}}
\newcommand{\cosmsinji}{\bm{cms}_{t,e,j,i}}

\newcommand{\vpJedan}{\bm{p}_{1,t,e,i,j}}
\newcommand{\vpDva}{\bm{p}_{2,t,e}}
\newcommand{\vpTri}{\bm{p}_{3,t,e,i,j}}

\newcommand{\condv}{\bm{\Lambda}_{t,e}}
\newcommand{\condc}{\bm{\Gamma}_{t,i,j}}
\newcommand{\condcji}{\bm{\Gamma}_{t,j,i}}
\newcommand{\conds}{\bm{\Phi}_{t,e,i,j}}

\newcommand{\qqbat}{p^{\mathrm{\es}}_{t}}
\newcommand{\qqch}{p^{\mathrm{ch}}_{t}}
\newcommand{\qqdis}{p^{\mathrm{dis}}_{t}}

\newcommand{\xxch}{x^{\mathrm{q}}_{t}}

\newcommand{\lambdaJedan}{\lambda_{t,i}}
\newcommand{\lambdaJedanDn}{\underline{\bm{\lambda}}_{1,t,i}}
\newcommand{\lambdaJedanUp}{\overline{\bm{\lambda}}_{1,t,i}}

\newcommand{\lambdaDva}{\lambda_{2,t,i}}
\newcommand{\lambdaDvaDn}{\underline{\bm{\lambda}}_{2,t,i}}
\newcommand{\lambdaDvaUp}{\overline{\bm{\lambda}}_{2,t,i}}

\newcommand{\lambdaTri}{\lambda_{3,t,e,i,j}}
\newcommand{\lambdaTriji}{\lambda_{3,t,e,j,i}}
\newcommand{\lambdaCetiri}{\lambda_{4,t,e,i,j}}
\newcommand{\lambdaCetiriji}{\lambda_{4,t,e,j,i}}
\newcommand{\lambdaPet}{\lambda_{5,t,e,i,j}}
\newcommand{\lambdaPetji}{\lambda_{5,t,e,j,i}}
\newcommand{\lambdaSest}{\lambda_{6,t,e,i,j}}
\newcommand{\lambdaSestji}{\lambda_{6,t,e,j,i}}
\newcommand{\lambdaSedam}{\lambda_{16,t,i}}

\newcommand{\lambdaJedanaest}{\lambda_{14,t,e,i,j}}
\newcommand{\lambdaDvanaest}{\lambda_{15,t,e,i,j}}
\newcommand{\lambdaTrinaest}{\mu_{5,t,e,i,j}}

\newcommand{\lambdaCetrnaest}{\lambda_{11,t,i,j}}
\newcommand{\lambdaPetnaest}{\lambda_{12,t,i,j}}
\newcommand{\lambdaSesnaest}{\mu_{2,t,i,j}}
\newcommand{\lambdaCetrnaestji}{\lambda_{11,t,j,i}}

\newcommand{\lambdaSedamnaest}{\lambda_{7,t,e,i,j}}
\newcommand{\lambdaOsamnaest}{\lambda_{8,t,e,i,j}}
\newcommand{\lambdaOsamnaestji}{\lambda_{8,t,e,j,i}}
\newcommand{\lambdaDevetnaest}{\lambda_{9,t,e,i,j}}
\newcommand{\lambdaDvadeset}{\mu_{1,t,e,i,j}}

\newcommand{\lambdaDvadesetjedan}{\lambda_{10,t,e,i,j}}
\newcommand{\lambdaDvadesetdva}{\lambda_{13,t,i,j}}

\newcommand{\muJedanDn}{\underline{\mu}_{3,t,k}}
\newcommand{\muJedanUp}{\overline{\mu}_{3,t,k}}
\newcommand{\muDvaDn}{\underline{\mu}_{4,t,k}}
\newcommand{\muDvaUp}{\overline{\mu}_{4,t,k}}
\newcommand{\muTriDn}{\underline{\mu}_{6,t,i}}
\newcommand{\muTriUp}{\overline{\mu}_{6,t,i}}

\newcommand{\cvJedan}{f_{1,t,i,j}}
\newcommand{\cvDva}{f_{2,t,i,j}}
\newcommand{\cvTri}{f_{0,t,i,j}}

\newcommand{\vvJedan}{w_{1,t,e,i,j}}
\newcommand{\vvDva}{w_{2,t,e,i,j}}
\newcommand{\vvTri}{w_{3,t,e,i,j}}
\newcommand{\vvCetiri}{w_{0,t,e,i,j}}

\newcommand{\Omegad}{\Omega^{\mathrm{d}}}
\newcommand{\Omegap}{\Omega^{\mathrm{p}}}

\newcommand{\xnula}{x_{0}}
\newcommand{\xjedan}{x_{1}}
\newcommand{\xdva}{x_{2}}
\newcommand{\xtri}{x_{3}}
\newcommand{\xcrta}{\overline{x}}
\newcommand{\x}{x}

\newcommand{\ynula}{y_{0}}
\newcommand{\yjedan}{y_{1}}
\newcommand{\ydva}{y_{2}}
\newcommand{\ytri}{y_{3}}
\newcommand{\ycrta}{\overline{y}}
\newcommand{\y}{y}

\newcommand{\eps}{\bm{\epsilon}}
\newcommand{\smf}{F}
\newcommand{\psin}{\psi_{n}}
\newcommand{\psijedan}{\psi_{1}}
\newcommand{\psidva}{\psi_{2}}
\newcommand{\un}{u_{n}}
\newcommand{\ujedan}{u_{1}}
\newcommand{\udva}{u_{2}}

\newcommand{\wbat}{w^{\mathrm{\es}}_{t}}
\newcommand{\wch}{w^{\mathrm{ch}}_{t}}
\newcommand{\wdis}{w^{\mathrm{dis}}_{t}}

\newcommand{\ybat}{y^{\mathrm{\es}}_{t}}

\newcommand{\ib}{b}
\newcommand{\iu}{u}
\newcommand{\xbin}{x^{\mathrm{bin}}_{t,\ib}}

\newcommand{\US}{U}
\newcommand{\BSS}{B}
\newcommand{\wbe}{w^{\mathrm{be}}_{t,\ib}}
\newcommand{\wue}{w^{\mathrm{ue}}_{t,\iu}}
\newcommand{\xun}{x^{\mathrm{ue}}_{t,\iu}}

\newcommand{\pen}{\bm{\pi}}

\newcommand{\Vr}{V_{t,i}^{\bm{r}}}
\newcommand{\Vi}{V_{t,i}^{\bm{i}}}

\newcommand{\Vrm}{V_{t,j}^{\bm{r}}}
\newcommand{\Vim}{V_{t,j}^{\bm{i}}}

\section{Appendix: Lower Level}\label{sec:App}
This section includes the formulation of the lower level, i.e. the exact AC OPF in the rectangular coordinates. Objective function \eqref{eq:A.1} minimizes production costs, \eqref{eq:A.2} and \eqref{eq:A.3} are bus balances, \eqref{eq:A.4}--\eqref{eq:A.7} are power flow equations, \eqref{eq:A.8} and \eqref{eq:A.9} are generator production limits, \eqref{eq:A.10} is line thermal limit and \eqref{eq:A.12} and \eqref{eq:A.13} are reference bus constraints. $\Vr$ and $\Vi$, $\tr$ and $\ti$ are real and imaginary parts of voltage magnitude and transformer tap ratio respectively. All other notations are the same as in our previous paper \cite{cpsota}.

\begin{equation}\tag{A.1}
\underset{\Xill}{\mathrm{Min}} \; \sum_{t,k}(\CC\cdot (\Pgen)^2 + \C \cdot \Pgen + \cc)
\label{eq:A.1}
\end{equation}

\begin{equation}\tag{A.2}
\begin{split}
&\sum_{k \in \BG}\Pgen - \sum_{l\in \BL}\dP -\hspace{-13pt} \sum_{(e,i,j)\in \A}\hspace{-13pt}\P - \underset{\hspace{-7pt}:i\in\BB}{\qbat}\\[4pt]
&- ((\Vr)^2+(\Vi)^2)\cdot\!\!\! \sum_{s\in \BS}\!\gs \!=\! 0, \!\!\quad \forall{t,i} \!\!\quad :\lambdaJedan
\label{eq:A.2}
\end{split}
\end{equation}

\begin{equation}\tag{A.3}
\begin{split}
&\sum_{k \in \BG}\Qgen - \sum_{l\in \BL}\dQ -\hspace{-13pt} \sum_{(e,i,j)\in \A}\hspace{-13pt}\Q\\[4pt]
&+ ((\Vr)^2+(\Vi)^2)\cdot\!\!\! \sum_{s\in \BS}\!\bs\! =\! 0, \!\!\quad \forall{t,i}  
\label{eq:A.3}
\end{split}
\end{equation}

\begin{equation}\tag{A.4}
\begin{split}
&\P= ((\Vr)^2+(\Vi)^2)\!\cdot\! (\gl+\gf)/\tap^2+(\\
&(-\gl \!\cdot\! \tr + \bl \!\cdot\! \ti)\!\cdot\!(\Vr\!\cdot\!\Vrm+\Vi\!\cdot\!\Vim) +\\
&(\bl \!\cdot\! \tr + \gl \!\cdot\! \ti) \!\cdot\!(\Vr\!\cdot\!\Vim-\Vi\!\cdot\!\Vrm))/\tap^2,\\
& \forall{t,(e,i,j)\in \AF}
\label{eq:A.4}
\end{split}
\end{equation}

\begin{equation}\tag{A.5}
\begin{split}
&\P= ((\Vr)^2+(\Vi)^2)\!\cdot\! (\gl+\gt)+(\\
&(-\gl \!\cdot\! \tr - \bl \!\cdot\! \ti)\!\cdot\!(\Vr\!\cdot\!\Vrm+\Vi\!\cdot\!\Vim) +\\
&(\bl \!\cdot\! \tr - \gl \!\cdot\! \ti) \!\cdot\!(\Vr\!\cdot\!\Vim-\Vi\!\cdot\!\Vrm))/\tap^2,\\
& \forall{t,(e,i,j)\in \AT}
\label{eq:A.5}
\end{split}
\end{equation}

\begin{equation}\tag{A.6}
\begin{split}
&\Q= -((\Vr)^2+(\Vi)^2)\!\cdot\! (\bl+\bf)/\tap^2+(\\
&(\bl \!\cdot\! \tr + \gl \!\cdot\! \ti)\!\cdot\!(\Vr\!\cdot\!\Vrm+\Vi\!\cdot\!\Vim) +\\
&(\gl \!\cdot\! \tr - \bl \!\cdot\! \ti) \!\cdot\!(\Vr\!\cdot\!\Vim-\Vi\!\cdot\!\Vrm))/\tap^2,\\
& \forall{t,(e,i,j)\in \AF}
\label{eq:A.6}
\end{split}
\end{equation}

\begin{equation}\tag{A.7}
\begin{split}
&\Q= -((\Vr)^2+(\Vi)^2)\!\cdot\! (\bl+\bf)/\tap^2+(\\
&(\bl \!\cdot\! \tr + \gl \!\cdot\! \ti)\!\cdot\!(\Vr\!\cdot\!\Vrm+\Vi\!\cdot\!\Vim) +\\
&(\gl \!\cdot\! \tr - \bl \!\cdot\! \ti) \!\cdot\!(\Vr\!\cdot\!\Vim-\Vi\!\cdot\!\Vrm))/\tap^2,\\
& \forall{t,(e,i,j)\in \AF}
\label{eq:A.7}
\end{split}
\end{equation}

\begin{equation}\tag{A.8}
\Pmin\! \le \Pgen \le \Pmax , \quad \forall{t,k}
\label{eq:A.8}
\end{equation}

\begin{equation}\tag{A.9}
\Qmin\! \le \Qgen \le \Qmax , \quad \forall{t,k}
\label{eq:A.9}
\end{equation}

\begin{equation}\tag{A.10}
\P^2\!+\!\Q^2 \! \le \Sline^2, \!\!\quad \forall{t,\!(e,i,j)\!\in\! \A } 
\label{eq:A.10}
\end{equation}

\begin{equation}\tag{A.11}
\Vmin^2 \le (\Vr)^2+(\Vi)^2 \le \Vmax^2, \quad \forall{t,i}
\label{eq:A.11}
\end{equation}

\begin{equation}\tag{A.12}
\Vr \ge 0, \quad \forall{t,i\in\RB}
\label{eq:A.12}
\end{equation}

\begin{equation}\tag{A.13}
\Vi = 0, \quad \forall{t,i\in\RB}
\label{eq:A.13}
\end{equation}

\bibliographystyle{IEEEtran}

\begin{thebibliography}{99}

\bibitem{ieee}
IEEE, ``IEEE Xplore''. Accessed: Oct. 4 2021. [Online]. Available: https://ieeexplore.ieee.org/Xplore/home.jsp.

\bibitem{terror}
J. M. Arroyo and F. D. Galiana, ``On the solution of the bilevel programming formulation of the terrorist threat problem,'' \emph{IEEE Transactions on Power Systems}, vol. 20, no. 2, pp. 789--797, May 2005.

\bibitem{pricing}
I. Momber, S. Wogrin and T. Gómez San Román, ``Retail Pricing: A Bilevel Program for PEV Aggregator Decisions Using Indirect Load Control,'' \emph{IEEE Transactions on Power Systems}, vol. 31, no. 1, pp. 464--473, Jan. 2016.

\bibitem{maintain}
C. Wang, Z. Wang, Y. Hou and K. Ma, ``Dynamic Game-Based Maintenance Scheduling of Integrated Electric and Natural Gas Grids With a Bilevel Approach,'' \emph{IEEE Transactions on Power Systems}, vol. 33, no. 5, pp. 4958--4971, Sept. 2018.

\bibitem{TEP}
L. P. Garces, A. J. Conejo, R. Garcia-Bertrand and R. Romero, ``A Bilevel Approach to Transmission Expansion Planning Within a Market Environment,'' \emph{IEEE Transactions on Power Systems}, vol. 24, no. 3, pp. 1513--1522, Aug. 2009.

\bibitem{energy_m}
E. Nasrolahpour, J. Kazempour, H. Zareipour and W. D. Rosehart, ``A Bilevel Model for Participation of a Storage System in Energy and Reserve Markets,'' \emph{IEEE Transasctions on Sustainable. Energy}, vol. 9, no. 2, pp. 582--598, April 2018.

\bibitem{fin_m}
M. Carrion, J. M. Arroyo and A. J. Conejo, ``A Bilevel Stochastic Programming Approach for Retailer Futures Market Trading,'' \emph{IEEE Transactions on Power Systems}, vol. 24, no. 3, pp. 1446--1456, Aug. 2009.

\bibitem{KKT}
S. A. Gabriel, A. J. Conejo, J. D. Fuller, B. F. Hobbs and C. Ruiz, ``Complementarity Modeling in Energy Markets,'' \textit{Springer}, 2013.

\bibitem{HPmaintain}
H. Pandžić, A. J. Conejo, I. Kuzle and E. Caro, ``Yearly Maintenance Scheduling of Transmission Lines Within a Market Environment,'' \emph{IEEE Transactions on Power Systems}, vol. 27, no. 1, pp. 407--415, Feb. 2012.


\bibitem{param}
K. Pand\v{z}i\'c, K. Bruninx and H. Pand\v{z}i\'c, ``Managing Risks Faced by Strategic Battery Storage in Joint Energy-Reserve Markets,'' \emph{IEEE Transactions on Power Systems}, vol. 36, no. 5, pp. 4355--4365, Sept. 2021.

\bibitem{cpsota}
K. Šepetanc and H. Pandžić, ``Convex Polar Second-Order Taylor Approximation of AC Power Flows: A Unit Commitment Study,'' \emph{IEEE Transactions on Power Systems}, vol. 36, no. 4, pp. 3585--3594, July 2021.

\bibitem{p1}
K. \v{S}epetanc, H. Pand\v{z}i\'c and T. Capuder, ``Solving Bilevel AC OPF Problems by Smoothing the Complementary Conditions -- Part I: Model Description and the Algorithm,'' \emph{Arxiv}, June 2022.

\bibitem{p2}
K. \v{S}epetanc, H. Pand\v{z}i\'c and T. Capuder, ``Solving Bilevel AC OPF Problems by Smoothing the Complementary Conditions -- Part II: Solution Techniques and Case Study,'' \emph{Arxiv}, June 2022.

\bibitem{KKTlin}
J. Fortuny-Amat and B. McCarl, ``A Representation and Economic Interpretation of a Two-Level Programming Problem,'' \emph{The Journal of the Operational Research Society}, vol. 32, no. 9, pp. 783--792 , Sept. 1981.

\bibitem{KKTquad}
T. A. Edmunds and J. F. Bard, ``Algorithms for nonlinear bilevel mathematical programs,'' \emph{IEEE Transactions on Systems, Man, and Cybernetics}, vol. 21, no. 1, pp. 83--89, Jan.--Feb. 1991.

\bibitem{discrete}
S. Zolfaghari and T. Akbari, ``Bilevel transmission expansion planning using second-order cone programming considering wind investment,'' \emph{Energy}, vol. 154, pp. 455--465 , July 2018.


\bibitem{BinInteraction}
B. Dandurand, K. Kim and S. Leyffer, ``A Bilevel Approach for Identifying the Worst Contingencies for Nonconvex Alternating Current Power Systems,'' \emph{SIAM Journal on Optimization}, vol. 31, no. 1, pp. 702-–726, Feb. 2021.

\bibitem{descent}
C. D. Kolstad and L. S. Lasdon, ``Derivative evaluation and computational experience with large bilevel mathematical programs,'' \emph{Journal of Optimization Theory and Applications}, vol. 65, no. 3, pp. 485-–499, June 1990.

\bibitem{SinglePen}
E. Aiyoshi and K. Shimizu, ``A solution method for the static constrained Stackelberg problem via penalty method,'' \emph{IEEE Transactions on Automatic Control}, vol. 29, no. 12, pp. 1111--1114, Dec. 1984.

\bibitem{BothPen}
Y. Ishizuka and E. Aiyoshi, ``Double penalty method for bilevel optimization problems,'' \emph{Annals of Operations Research}, vol. 34, no. 1, pp. 73--88, Dec. 1992.

\bibitem{QPTrust}
B. Colson, P. Marcotte and G. Savard, ``A Trust-Region Method for Nonlinear Bilevel Programming: Algorithm and Computational Experience,'' \emph{Computational Optimization and Applications}, vol. 30, no. 3, pp. 211--227, March 2005.

\bibitem{param1}
H. C. Bylling, S. A. Gabriel and T. K. Boomsma, ``A parametric programming approach to bilevel optimisation with lower-level variables in the upper level,'' \emph{Journal of the Operational Research Society}, vol. 71, no. 5, pp. 846--865, 2020.

\bibitem{param2}
N. P. Faísca, V. Dua, B. Rustem, P. M. Saraiva and E. N. Pistikopoulos, ``Parametric global optimisation for bilevel programming,'' \emph{Journal of Global Optimization}, vol. 38, pp. 609–-623, Aug. 2007.

\bibitem{param3}
S. Avraamidou and E. N. Pistikopoulos, ``A Multi-Parametric optimization approach for bilevel mixed-integer linear and quadratic programming problems,'' \emph{Computers \& Chemical Engineering}, vol. 125, pp. 98–-113, June 2019.

\bibitem{nested}
A. Sinha, P. Malo, A. Frantsev and K. Deb, ``Finding optimal strategies in a multi-period multi-leader–follower Stackelberg game using an evolutionary algorithm,'' \emph{Computers \& Operations Research}, vol. 41, pp. 374--385, Jan. 2014.

\bibitem{PSO}
X. Li, P. Tian, and X. Min, ``A Hierarchical Particle Swarm Optimization for Solving Bilevel Programming Problems,'' \emph{Artificial Intelligence and Soft Computing -- ICAISC 2006} (LNCS 4029), L. Rutkowski, R. Tadeusiewicz, L. A. Zadeh and J. M. Zurada, Eds. Heidelberg, Germany: Springer, 2006, pp. 1169—-1178.

\bibitem{DE}
J. S. Angelo, E. Krempser and H. J. C. Barbosa, ``Differential evolution for bilevel programming,'' \emph{2013 IEEE Congress on Evolutionary Computation}, 2013, pp. 470--477.

\bibitem{hibridDE}
X. Zhu, Q. Yu and X. Wang, ``A Hybrid Differential Evolution Algorithm for Solving Nonlinear Bilevel Programming with Linear Constraints,'' \emph{2006 5th IEEE International Conference on Cognitive Informatics}, 2006, pp. 126--131.

\bibitem{EvolutionKKT}
S. R. Hejazi, A. Memariani, G. Jahanshahloo and M. M. Sepehri, ``Linear bilevel programming solution by genetic algorithm,'' \emph{Computers \& Operations Research}, vol. 29, no. 13, pp. 1913--1925, Nov. 2002.

\bibitem{OptimalSolutionMap}
A. Sinha, P. Malo and K. Deb, ``Evolutionary algorithm for bilevel optimization using approximations of the lower level optimal solution mapping,'' \emph{European Journal of Operational Research}, vol. 257, no. 2, pp. 395--411, March 2017.

\bibitem{OptValueFun}
A. Sinha, P. Malo and K. Deb, ``Solving optimistic bilevel programs by iteratively approximating lower level optimal value function,'' \emph{2016 IEEE Congress on Evolutionary Computation (CEC)}, 2016, pp. 1877--1884.

\bibitem{tutorial}
A. Sinha, P. Malo and K. Deb, ``A Review on Bilevel Optimization: From Classical to Evolutionary Approaches and Applications,'' \emph{IEEE Transactions on Evolutionary Computation}, vol. 22, no. 2, pp. 276--295, Apr. 2018.

\bibitem{Knjiga1}
A. Ivakhnenko and V. Lapa, \emph{Cybernetic predicting devices}, N.Y.: CCM Information Corp., 1973.

\bibitem{Knjiga2}
A. Ivakhnenko and V. Lapa, \emph{Cybernetics and forecasting techniques}, N.Y.: American Elsevier, 1967.

\bibitem{DC}
B. Stott, J. Jardim, and O. Alsac, ``DC Power Flow Revisited,`` \emph{IEEE Trans. Power Syst.}, vol. 24, no. 3, pp. 1290--1300, Aug. 2009.

\bibitem{eem}
H. Pandžić and I. Kuzle, ``Energy storage operation in the day-ahead electricity market,'' \emph{2015 12th International Conference on the European Energy Market (EEM)}, 2015, pp. 1--6.

\bibitem{relax}
B. Dandurand, K. Kim and S. Leyffer, ``A Bilevel Approach for Identifying the Worst Contingencies for Nonconvex Alternating Current Power Systems,'' \emph{SIAM Journal on Optimization}, vol. 31, no. 1, pp. 702-–726, Feb. 2021.

\bibitem{relax2}
S. Zolfaghari and T. Akbari, ``Bilevel transmission expansion planning using second-order cone programming considering wind investment,'' \emph{Energy}, vol. 154, pp. 455-–465, April 2018.

\bibitem{basic_ACOPF}
S. Frank and S. Rebennack, ``An introduction to optimal power flow: Theory, formulation, and examples,'' \emph{IIE Transactions}, vol. 48, no. 12, pp. 1172--1197, May 2016.

\bibitem{IV}
R. P. O’Neill, A. Castillo and M. B. Cain, ``The IV formulation and linear approximations of the ac optimal power flow problem,'' FERC, Washington, DC, USA, Tech. Rep., Dec. 2012. Accessed on: Jul. 10, 2021. [Online] Available at: \url{cms.ferc.gov/sites/default/files/2020-05/acopf-2-iv-linearization.pdf}

\bibitem{D1}
K. Hornik, M. Stinchcombe and H. White, ``Multilayer feedforward networks are universal approximators,'' \emph{Neural Networks}, vol. 2, no. 5, pp. 359--366, Mar. 1989.

\bibitem{D2}
G. Cybenko, ``Approximation by superpositions of a sigmoidal function,'' \emph{Mathematics of Control, Signals, and Systems}, vol. 2, no. 4, pp. 303--314, Dec. 1989.

\bibitem{D3}
K. Fukushima, ``Neocognitron: A self-organizing neural network model for a mechanism of pattern recognition unaffected by shift in position,'' \emph{Biological Cybernetics}, vol. 36, no. 4, pp. 193--202, Apr. 1980.

\bibitem{D4}
D. Ciregan, U. Meier and J. Schmidhuber, ``Multi-column deep neural networks for image classification,'' \emph{2012 IEEE Conference on Computer Vision and Pattern Recognition}, 2012, pp. 3642--3649.

\bibitem{D5}
J. Schmidhuber, ``Deep learning in neural networks: An overview,'' \emph{Neural Networks}, vol. 61, pp. 85--117, Jan. 2015.

\bibitem{Nature}
D. Rumelhart, G. Hinton and R. Williams, ``Learning representations by back-propagating errors,'' \emph{Nature}, vol. 323, no. 6088, pp. 533-536, Oct. 1986.

\bibitem{AD}
R. B. Rall, \emph{Automatic Differentiation: Techniques and Applications}, N.Y: Springer-Verlag, 1981.

\bibitem{D6}
A. Paszke \emph{et al.}, ``PyTorch: An Imperative Style, High-Performance Deep Learning Library,'' \emph{Advances in Neural Information Processing Systems 32}, 2019, pp. 1--12.

\bibitem{fastai}
J. Howard \emph{et al.} \emph{fastai}, GitHub, 2018. Accessed on: Aug. 17, 2021. [Online] Available at: \url{github.com/fastai/fastai}

\bibitem{D7}
D. P. Kingma and J. Ba, ``Adam: A Method for Stochastic Optimization,'' \emph{3rd International Conference for Learning Representations}, 2015, San Diego, USA.

\bibitem{DD}
T. Ivek and D. Vlah, ``BlackBox: Generalizable reconstruction of extremal values from incomplete spatio-temporal data,'' \emph{Extremes}, vol. 24, no. 1, pp. 145-–162, Oct. 2020.

\bibitem{Ranger}
L. Wright \emph{et al.} \emph{Ranger Deep Learning Optimizer}, GitHub, 2019. Accessed on: Aug. 17, 2021. [Online] Available at: \url{https://github.com/lessw2020/Ranger-Deep-Learning-Optimizer}

\bibitem{D8}
L. Liu, H. Jiang, P. He, W. Chen, X. Liu, J. Gao and J. Han, ``On the Variance of the Adaptive Learning Rate and Beyond,'' \emph{2020 International Conference on Learning Representations}, 2020.

\bibitem{D9}
M. R. Zhang, J. Lucas, G. Hinton and J. Ba, ``Lookahead Optimizer: k steps forward, 1 step back,'' \emph{Neural Information Processing Systems 2019}, 2019.

\bibitem{D10}
L. N. Smith, ``A disciplined approach to neural network hyper-parameters: Part 1 - learning rate, batch size, momentum, and weight decay,'' \emph{arXiv}, 2018.

\bibitem{pglib}
C. Coffrin \emph{et al.} \emph{PGLib-OPF v20.07}, GitHub, Jul. 29, 2020. Accessed on: Sept. 15, 2021. [Online] Available at: \url{github.com/power-grid-lib/pglib-opf/tree/v20.07}

\bibitem{rts96}
C. Grigg \emph{et al}., ``The IEEE Reliability Test System-1996,'' \emph{IEEE Trans. Power Syst.}, vol. 14, no. 3, pp. 1010--1020, Aug. 1999.

\bibitem{naja}
K. Pand\v{z}i\'c, H. Pand\v{z}i\'c and I. Kuzle, ``Coordination of Regulated and Merchant Energy Storage Investments,'' \emph{IEEE Transactions on Sustainable Energy}, vol. 9, no. 3, pp. 1244--1254, June 2022.





\end{thebibliography}

\end{document}